# Which way up? Recognition of homologous DNA segments in parallel and antiparallel alignment

Dominic J. (O') Lee[1,a.)], Aaron Wynveen[2], Tim Albrecht[1], Alexei A. Kornyshev[1,b.)]

[1] *Department of Chemistry, Imperial College London, SW7 2AZ, London, UK*

[2] *School of Physics and Astronomy, University of Minnesota, Minneapolis, Minnesota 55455, USA*

## Abstract

Homologous gene shuffling between DNA molecules promotes genetic diversity and is an important pathway for DNA repair. For this to occur, homologous genes need to find and recognize each other. However, despite its central role in homologous recombination, the mechanism of homology recognition has remained an unsolved puzzle of molecular biology. While specific proteins are known to play a role at later stages of recombination, an initial coarse grained recognition step has been proposed. This relies on the sequence dependence of the DNA structural parameters, such as twist and rise, mediated by intermolecular interactions, in particular electrostatic ones. In this proposed mechanism, sequences that have the same base pair text, or are homologous, have lower interaction energy than those sequences with uncorrelated base pair texts. The difference between the two energies is termed the 'recognition energy'. Here, we probe how the recognition energy changes when one DNA fragment slides past another, and consider, for the first time, homologous sequences in antiparallel alignment. This dependence on sliding was termed the 'recognition well'. We find that there is recognition well for anti-parallel, homologous DNA tracts, but only a very shallow one, so that their interaction will differ little from the interaction between two nonhomologous tracts. This fact may be utilized in single molecule experiments specially targeted to test the theory. As well as this, we test previous theoretical approximations in calculating the recognition well for parallel molecules against MC simulations, and consider more rigorously the optimization of the orientations of the fragments about their long axes upon calculating these recognition energies. The more rigorous treatment affects the recognition energy a little, when the molecules are considered rigid. However when torsional flexibility of the DNA molecules is introduced, we find excellent agreement between the analytical approximation and simulations.

---

[a.)] Electronic Mail: domolee@hotmail.com
[b.)] Electronic Mail: domolee@hotmail.com

# 1. Introduction

The shuffling of homologous genes between DNA molecules is a key process in meiosis and DNA repair. It expedites evolution and facilitates genetic diversity, as well as underpinning the transfer of genetic material between different strains and species of bacteria and viruses. The recombination process requires base pairing between the recombining strands, in which various proteins, for instance the RecA protein in E-coli, play a crucial role. However, before recombination there needs to be a search in which homologous genes can find each other. The slow process of random diffusion is incompatible with the rate of recombination [1,2], suggesting that another mechanism is at play. What this might be is an important remaining question in molecular biology.

The process of homologous recombination is known to proceed via (i) the breakage of double *s*tranded (ds-)DNA and formation of single strands mediated by specialized proteins [3,4], followed by (ii) the single strand recognizing and invading a homologous double helix through base pairing (single strand invasion). One possible stage where the recognition of the homologous genes takes place is at the stage of broken strand exchange, utilizing the sequence complementarity between single strands [4,5,6,7]. However, if complementarity was the only homology recognition mechanism at work, recognition would occur between homologous fragments of genes with as few as 8-10 base pairs [8], suggesting that frequent recognition and recombination errors would be inevitable. It might also take too long for the single strand to search for the homologous duplex. Indeed, there is evidence to suggest that before strand breakage occurs, identical intact double-stranded DNA segments may pair [1,9,10,11,12]. It is also conceivable that homology recognition between intact DNA tracts may occur in other biological processes. One instance might be the silencing of multiple copies of the same gene [10]. One possible mechanism for homology recognition before double strand breakage is the stem loop kissing model [13,14,15,16], which again relies on the sequence complementarity of single stranded DNA, at loop ends. A second alternative for homology pairing may arise from the base-pair dependent conformation of DNA [17,18].

It has been well established that DNA wraps on histones at defined sequence tracts [19], the base sequence simultaneously encoding the nucleosome positioning [20]. This may arise from base-pair sequence dependence in the bending of DNA [21,22], but also from the base pair dependence of DNA structure [23,24]. Base pair dependent DNA structure may also influence the interactions between two DNA fragments in modes of interaction that depend on the helix shape of the molecule [25,26, 27,28]. There is some compelling evidence to suggest that the intermolecular interactions are indeed affected by the helix structure of DNA [29,30]. If the helix-dependent forces are important then a base pair specific pattern of helix distortions [31] may well provide a mechanism for homology recognition [32]. This relies on the fact that two paired DNA fragments with the same pattern of helix distortions have lower interaction energy than those with different patterns [32,33,34,35] due to commensurability between positive counter-ion and negative phosphate charges along the DNA pair, known as an electrostatic zipper [36].

Several experimental studies have reported signatures of DNA-DNA recognition, at the molecular level, in protein free solutions. These suggest a possible role for direct interactions between double stranded DNA in the homologous pairing process. The experiments utilized gel electrophoresis [37]; cholesteric liquid crystals [38]; AFM imaging [39]; and surface-confined DNA using magnetic beads [40]. In addition, Seeman's group reported the formation of PX-DNA complexes presumably resulting from homologue pairing [41], and homology recognition has also been observed between nucleosomes [39].

In two recent publications [42,43], in which the forces that depend on helix structure were considered, the effect of sliding one molecule with respect to another on interaction energies between homologous sequences, aligned in the same direction, was investigated. In Ref. [42] the DNA was considered to be rigid, whereas in Ref. [43] torsional elasticity was included. These studies demonstrated that the interaction energy, as a function of the sliding distance, forms a potential well centred at the point where the two homologous base pair sequences are perfectly aligned in parallel register. The width of the well has been found to be much larger than the decay length of DNA-DNA interactions (a Debye screening length of the order of Ångstroms), contrary to naive expectations. Its origin lies in the character of the helix distortions [42,43]. This finding advocates a possible role for helix structure dependent forces in aligning homologous sequences before single strand invasion. Indeed, no energy well was predicted for non-homologous molecules. When understanding the pairing mechanism, as in Refs. [32,42,43], the absence of a recognition well between homologous sequences aligned in opposite directions (antiparallel homology) seems intuitively obvious, and was suggested in Ref. [42] without rigorous analysis.

In this study, we test the theoretical results of the previous work [42,43] against Monte-Carlo simulations. We also rigorously investigate, for the first time, the pairing energy for DNA pairs containing antiparallel homology. One finding is that there is, in fact, a recognition well, but a very shallow one. As the length of the antiparallel homologous sequences in juxtaposition increases, the difference in the pairing energy per base pair between such pairs and those containing totally uncorrelated sequences becomes negligible. Another issue that this study raised, for rigid molecules, is how the optimization of the interaction energy between two molecules with respect to the angles of rotation of the DNA fragments about their long axes should be performed. In previous studies, the optimization was performed after averaging over all realizations of base-pair sequence-dependent distortions of the two DNA fragments. For rigid molecules, the sequence dependent base pair disorder is quenched, which is to say that it is frozen and cannot change, unlike thermal disorder. Thus, strictly speaking, one should optimise the interaction energy for each possible set of base pair sequences first, before averaging the energies of an ensemble of different sequences. In the present study we have investigated the impact of the order of averaging on the evaluated recognition energy (the pairing energy difference between DNA fragments containing homologous sequences and those that do not). We find that the recognition energy changes only slightly; moreover, when we introduce torsional flexibility for long homologous sequences, this effect vanishes.

The paper is structured in the following way. In the theory section, we discuss the nature of paired fragments investigated, as well as the model of helix distortions. Then, we consider the case of rigid molecules, where we present the two ways of optimizing the azimuthal orientations of the two fragments about their long-axes. Lastly, we describe the situation of torsionally flexible molecules. The results section is divided into two parts. In the first we consider rigid molecules and, in the second, torsionally flexible molecules. In the discussion section we discuss our findings and point to future theoretical and experimental work. Lastly, we present a summary of the key findings of this study.

## 2. Theory

*General Considerations*

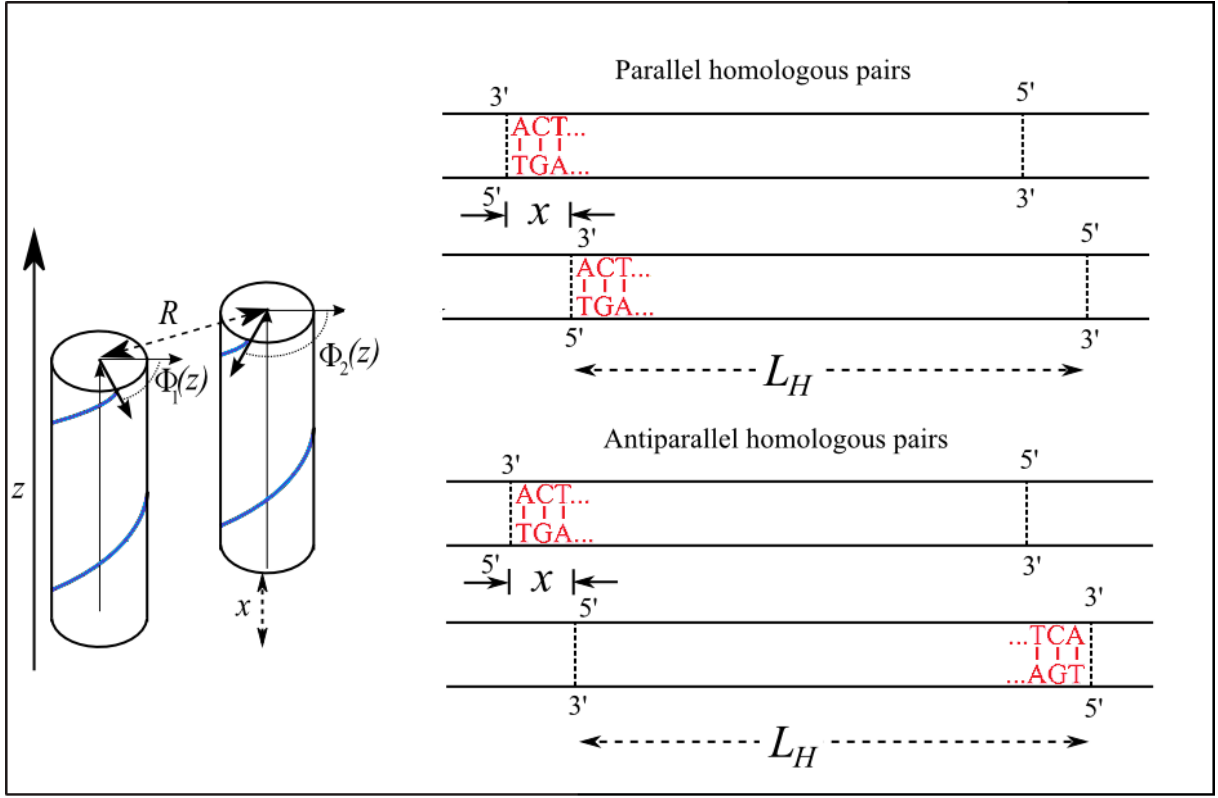

Fig.1. Diagrams illustrating the orientations of the paired fragments. On the left hand side, we show a small section of any of the three types of paired DNA fragments. Here, the trajectory of the point where each minor groove is bisected is shown in blue. The z- axis runs parallel to the molecular centre lines, illustrated by grey lines. The equation describing these trajectories is given by Eq. (1). The two DNA fragments are separated by interaxial distance $R$ and the respective azimuthal orientations of each trajectory are described by the phases $\Phi_1(z) + \phi_0 = \delta\Phi_1(z) + gz$ and $\Phi_2(z) + \phi_0 = \delta\Phi_2(z) + zg$. Here, $\phi_0$ is a constant chosen here for the ease of presentation. It is set so that the points of reference from which $\Phi_1(z)$ and $\Phi_2(z)$ are measured do not lie on the line connecting the two molecular centre lines. The phase difference, which interaction energy depends on, is given by $\Delta\Phi(z) = \delta\Phi_1(z) - \delta\Phi_2(z) = \Phi_1(z) - \Phi_2(z)$, and does not depend on this choice. On the right hand side, we illustrate what we mean by antiparallel and parallel aligned homologies. Here, we present long DNA fragments of arbitrary sequences, which have homologous (identical) segments; parts of the arbitrary base pair text are shown. $L_H$ is the length of homology (the unshown length $L$ corresponds to the whole fragment); the short dashed lines refer to the ends of the homologous segments. Outside these lines, the sequences are uncorrelated between the two fragments. The end of one homology fragment may be shifted a distance $x$ away from the end of the second fragment. Positions of both the 3' and 5' ends of the homologous segments define the direction of DNA polarity. Indeed, we could define a different antiparallel

homology where we could swap the 3' and 5' ends. This would be chemically distinct from the previous case in which we just rotated one segment by 180° from the parallel orientation; because of this, the pattern of distortions in the 'swapped' case could be different.

To investigate DNA homology recognition through helix dependent interactions, we construct a model in which two DNA fragments, each with total length $L$, lie parallel to each other, along their long axis ($z$-axis). Three different types of paired DNA fragments are considered. First, we suppose that the two fragments have completely different uncorrelated base pair texts, referring to this as a random pair. The other two types of pairing consist of tracts of DNA which have the same text (or are homologous to each other [44]), and are of length $L_H$, which, in turn, are embedded into fragments with random pairing. These final two are distinguished by the homologous tracts being either aligned so that both texts read in the same direction (*parallel homology pairing*) or in the opposite direction (*antiparallel homology pairing*). The homologous sequences are positioned with their centres coinciding with the centres of both fragments. Therefore, on either side we can have tracts of DNA of length $L_{NH} = (L - L_H)/2$, which consist of sequences which are completely uncorrelated with respect to each other. We label the two fragments within a pair $\mu = 1, 2$. This arrangement is illustrated in Fig.1. The case where $L_{NH} \neq 0$ is not necessarily relevant to the recombination of genes. Nevertheless, we consider it for two reasons. First, this is the situation in homology recognition experiments currently underway. Secondly, this situation could be realized in other biological processes that may require the pairing of identical sequences, for instance repeat induced point mutation [45,46].

To probe how the homologous segments recognize each other, we shift the centre of fragment 2 a distance $x$ along the z-axis. Therefore, the centre of fragment 1 lies at $z = L/2$, whereas that of fragment 2 lies at $z = L/2 + x$. For each pair we first calculate the pairing energy, either at $x = 0$ or as a function of $x$. For rigid fragments the pairing energy is simply the interaction energy, but for the pairing energy of torsionally flexible molecules, elastic energy will also play a role (see below). Upon determining these energies we calculate the difference, or 'recognition energy', between the interaction energies of fragments with embedded homologous sequences and the energies of pairs containing completely uncorrelated, nonhomolgous, sequences.

In the model that we consider for homology recognition, differences in pairing energy arise from DNA not being an ideal helix structure. Previous studies [31] have shown that a particular base pair text leads to its own pattern of helix distortions. To understand what is meant by helix distortions, let us consider the trajectories that the minor grooves trace out along the two fragments. First, we suppose that the molecular axes of the two fragments are separated a distance $R$ apart. Then, the location of the minor groove along the helices, with radius $a$, can be described through the position vectors

$$\mathbf{r}_\mu(z) = \left(\frac{(-1)^\mu R}{2} + a\cos(\delta\Phi_\mu(z) + gz)\right)\hat{\mathbf{i}} + a\sin(\delta\Phi_\mu(z) + gz)\hat{\mathbf{j}} + z\hat{\mathbf{k}}, \qquad (1)$$

in the region $0 < z < L$, for fragment 1, and $x < z < L + x$ for fragment 2. The trajectories described by Eq. (1) are illustrated on the left hand side of Fig. 1. Here, $g = 2\pi / H$, where $H \approx 33.8\text{Å}$ is the average helical pitch of DNA. If the phase $\delta\Phi_\mu(z)$ is constant (or linear) with respect to $z$, Eq. (1) defines an ideal helix. It is the variations in $\delta\Phi_\mu(z)$ along the helices that describe the pattern of distortions for each fragment. These variations arise from both thermal fluctuations and intrinsic base-pair-specific distortions [25,31,47].

When there is no torsional strain on the two molecules (the torsionally relaxed state)

$$\frac{d\delta\Phi_\mu(z)}{dz} = \frac{\delta\Omega_\mu(z)}{h}. \tag{2}$$

The function $\delta\Omega_\mu(z)$ represents the field that characterizes the pattern of base pair distortions of an isolated molecule without thermal fluctuations. A particular function $\delta\Omega_\mu(z)$ representing a particular base pair sequence can be constructed from available DNA structural information [31]. We assume that (for justification see Ref. [31])

$$\left\langle \delta\Omega_\mu(z)\delta\Omega_\mu(z') \right\rangle_{\delta\Omega} = \frac{h^2}{\lambda_c^{(0)}} \delta(z - z'), \tag{3}$$

where the bracket denotes an ensemble average over all possible realizations of $\delta\Omega_\mu(z)$. Here, $h \approx 3.4\text{Å}$ is the average distance (or rise) between base pairs and $\lambda_c^{(0)}$ is the structural contribution to what we term the coherence length [32,31,25], which we take to be $\lambda_c^{(0)} = 150\text{Å}$ based on previous estimates [31]. The parameter $\lambda_c^{(0)}$ is a measure of the degree of helix non-ideality caused by base-pair dependent distortions. It is the length scale over which two distorted helices with different base pair texts, described through Eqs. (1)-(3), lose alignment with each other [25, 47].

For random pairs, whose distortions are uncorrelated, $\left\langle \delta\Omega_1(z)\delta\Omega_2(z') \right\rangle = 0$. For the other two pairing types, the random sequences on either side of the homologous segments are also uncorrelated with each other. Thus, as before, $\left\langle \delta\Omega_1(z)\delta\Omega_2(z') \right\rangle = 0$ when either $0 < z, z' - x < L_{NH}$ or $L_{NH} + L_H < z, z' - x < 2L_{NH} + L_H$. When $L_{NH} < z < L_{NH} + L_H$, for parallel homologous pairs $\delta\Omega_1(z) = \delta\Omega_2(z - x)$, whereas for the antiparallel pairs $\delta\Omega_1(z) = \delta\Omega_2(L_H + 2L_{NH} - z + x)$. In the main text we will consider only the case where $L_{NH} = 0$. In the Supplemental Material we show how the recognition energies depend on $L_{NH}$ for rigid molecules. Any effect due to non-homologous ends becomes negligible when we introduce torsional flexibility for sufficiently long homologous segments, as seen in the Supplemental Material.

If the interaction energy $E_{int}$ depends on the helix structure of the DNA, we may write [25,32]:

$$E_{int} = La_0(R) + \int_x^L dz \left( -a_1(R) \cos \Delta\Phi(z) + a_2(R) \cos 2\Delta\Phi(z) \right), \tag{4}$$

where $\Delta\Phi(z) = \delta\Phi_1(z) - \delta\Phi_2(z)$, is the difference between the sequence dependent distortions of paired molecules. In what follows, we will consider two separate situations. First, as in Ref. [42], the molecules are considered as torsionally rigid such that no torsional strain can accumulate. Hence, the relative angle between base-pairs on opposing molecules is solely due to sequence-dependent variations, i.e., Eq. (2). The second case is when we allow for torsional flexibility. Here, changes in $\Delta\Phi(z)$ from the form specified by Eq. (2) may occur in order to reduce (optimize) the interaction energy (described by Eq. (4)) at the cost of torsional elastic energy. In the main text we shall consider only the ground state, leaving the effect of thermal fluctuations to the Supplemental Material.

*Rigid Molecules*

For the case of rigid molecules, we consider only the $x = 0$ case, where, for homologous segments, the sequence-dependent distortions of paired molecules lie directly across from one another. One can easily integrate Eq. (2), which yields

$$\Delta\Phi(z) = \frac{1}{h} \int_0^z \left[ \delta\Omega_1(z') - \delta\Omega_2(z') \right] dz' + \Delta\Phi_0, \tag{5}$$

where $\Delta\Phi_0$ is a constant of integration to be determined by minimizing the interaction energy either after or before averaging over the realizations of $\delta\Omega_\mu(z')$, the sequence-dependent distortions describing a particular DNA sequence. $\Delta\Phi_0$, in geometric terms, is the azimuthal angle between the lines bisecting each minor groove of the two molecules at the position $z = 0$. In previous work, Ref. [42], $\Delta\Phi_0$ was determined after averaging. In this case, $\Delta\Phi_0$ takes the form

$$\Delta\Phi_0 = \Delta\bar{\Phi}_0 - \frac{1}{h} \int_0^d \left[ \delta\Omega_1(z') - \delta\Omega_2(z') \right] dz'. \tag{6}$$

Eq. (6) ensures that at the location $d$, $\Delta\Phi(d) = \Delta\bar{\Phi}_0 = \langle \Delta\Phi(z) \rangle_{\delta\Omega}$. In other words, at $z = d$, $\Delta\Phi(z)$ is fixed at its ensemble-averaged value. The ensemble-averaged value $\Delta\bar{\Phi}_0$ is determined by both the values $a_1(R)$ and $a_2(R)$ in the interaction energy (Eq. (4)). By substituting Eq. (6) into Eq. (5), we may obtain general analytical forms for the average pairing energies for our three types of pairing $\langle E_{int} \rangle_{\delta\Omega}$, using the appropriate forms of $\delta\Omega_\mu(z')$ for each (see Appendix A of the Supplemental Material). In in this study, we have chosen that $a_1(R) > 4a_2(R)$, so that $\Delta\bar{\Phi}_0 = 0$. Here, we find that the minimum average energy with respect to $d$ is given at the location $d = L/2$ for both fragments containing parallel homologous segments and random pairs. For antiparallel fragments, there are two optimal values of $d$ at $d = L_H/2 \pm L_H/4$ (for $L_{NH} = 0$). For these cases, we find the following expressions for the two recognition energies (the energy difference between the homologous pairs and uncorrelated ones), for details see Appendix A of the Supplemental Material. For parallel homologous pairs we have ($L_{NH} = 0$)

$$E_{rec} = -L_H \left(a_1(R) - a_2(R)\right) + 2\lambda_c^{(0)} \left[ a_1(R)\left(1 - \exp\left(-\frac{L_H}{2\lambda_c^{(0)}}\right)\right) - \frac{a_2(R)}{4}\left(1 - \exp\left(-\frac{2L_H}{\lambda_c^{(0)}}\right)\right) \right],$$

(7)

and for antiparallel homologous pairs ($L_{NH} = 0$)

$$E_{rec} = \left\{ -a_1(R)\lambda_c^{(0)} \left[ 2 - 4\exp\left(-\frac{L_H}{4\lambda_c^{(0)}}\right) + 2\exp\left(-\frac{L_H}{2\lambda_c^{(0)}}\right) \right] \right.$$
$$\left. + a_2(R)\lambda_c^{(0)} \left[ \frac{1}{2} - \exp\left(-\frac{L_H}{\lambda_c^{(0)}}\right) + \frac{1}{2}\exp\left(-\frac{2L_H}{\lambda_c^{(0)}}\right) \right] \right\}.$$

(8)

We also performed MC simulations over an ensemble of fragments with different sequence-dependent distortions. For some of these simulations, $\Delta\Phi_0$ is found after averaging over all the realizations of $\delta\Omega_1(z)$ and $\delta\Omega_2(z)$. These simulations have been compared to the analytical results given by Eqs.(7) and (8), yielding excellent agreement (c.f. Fig.2).

Considering that we are dealing with disorder that is fixed and cannot change (quenched disorder), unlike thermal fluctuations, a better way of choosing $\Delta\Phi_0$ is to minimize the interaction energy of a particular configuration described by $\delta\Omega_\mu(z)$ and taking the average of the optimized energies for these configurations, rather than ensemble averaging over these configurations and then optimizing over $\Delta\Phi_0$ to find the minimum energy of this ensemble average. For the former case, on inspection, it is an extremely laborious task to obtain the exact analytical solution for $\Delta\Phi_0$ in terms of the arbitrary function $\delta\Omega_1(z') - \delta\Omega_2(z')$, and it does not seem possible to obtain a formula for the average energy if one had such a solution. However, we can perform MC simulations that directly minimize $\Delta\Phi_0$ for each configuration of $\delta\Omega_1(z') - \delta\Omega_2(z')$ in order to determine the ensemble average of the interaction energy. In addition, we have obtained the analytical result of the leading order term in a perturbation expansion in $\delta\Omega_1(z') - \delta\Omega_2(z')$. Such an expansion assumes that $L$ is not too large compared to $\lambda_c$. Details on how such a perturbation expansion is performed are given in Appendix B of the Supplemental Material. To leading order we find that

$$\Delta\Phi_0 \approx \Delta\bar{\Phi}_0 - \frac{1}{hL}\int_0^L dz' \int_0^s dz'' \left[\delta\Omega_1(z'') - \delta\Omega_2(z'')\right].$$ (9)

Eq. (9) essentially supposes that the spatial average of $\Delta\Phi(z)$ is $\Delta\bar{\Phi}_0 = \langle\Delta\Phi(z)\rangle_{\delta\Omega}$. Again, expressions for the recognition energies may be derived by substituting Eq. (9) into Eq. (4). We have refrained from giving these expressions in the main text since they are rather cumbersome and such an approximate theory does not yield good quantitative results upon comparing them to the approximation-free MC –results that we have obtained. However, the theory provides some qualitative insight into what happens when one changes $L_{NH}$ [48]. Also the perturbation result, which correctly reproduces the limiting behaviour for small $L$, could be used (as discussed later) as the basis of a more comprehensive theory, extending over the full range of $L_{NH}$ and $L_H$ and including torsional flexibility.

*Torsionally flexible molecules*

For torsionally flexible molecules, we consider the elastic rod model and introduce a torsional elastic term [43]. The total pairing energy becomes

$$E_{pair} = \frac{C}{4} \int_x^L dz \left( \frac{d\Delta\Phi(z)}{dz} - \frac{\delta\Omega_1(z) - \delta\Omega_2(z)}{h} \right)^2 + E_{int}, \tag{10}$$

where $C$ is the torsional elasticity of the molecules. For finite $C$, Eq. (2) is no longer obeyed. In fact, upon minimizing the energies given by the combination of Eqs. (4) and (10), $\Delta\Phi(z)$ now satisfies a sine –Gordon like non-linear equation [32], which has an inhomogeneous term that depends on the derivative of $\delta\Omega_1(z) - \delta\Omega_2(z)$ with respect to $z$ (for details see Ref. [33] or [34]). Unfortunately, an exact analytical solution to this equation in terms of $\delta\Omega_1(z) - \delta\Omega_2(z)$ cannot be obtained, although some progress has been made by using path integral methods (see Appendix C of Ref. [49]). Here, we use a much simpler approach. If the helix dependent interactions described by Eq. (4) are sufficiently strong, we may use a trial function of the form (originally suggested in Ref. [33])

$$\Delta\Phi = \Delta\bar{\Phi}_0 + \frac{1}{2h} \int_{-\infty}^{\infty} \text{sgn}(z-z') \exp\left(-\frac{|z-z'|}{\lambda_h}\right) [\delta\Omega_1(z') - \delta\Omega_2(z')] dz' \tag{11}$$

and substitute this into Eqs. (4) and (10), performing the average over realizations of $\delta\Omega_1(z') - \delta\Omega_2(z')$. The parameters $\lambda_h$ and $\Delta\bar{\Phi}_0$ are chosen to minimize the average energy. The parameter $\lambda_h$ is a correlation length, which we term the adaptation length [33, 47]. Its value increases as we reduce the strength of helix dependent interactions, or if we increase the size of torsional rigidity $C$. Here, for parallel homologous pairs we choose $\delta\Omega_1(z) = \delta\Omega_2(z-x)$ and for antiparallel homologous pairs $\delta\Omega_1(z) = \delta\Omega_2(x-z)$, and use Eq. (11) exclusively to describe the homologous regions. This approach was used in Ref. [43] for sliding fragments ($x \neq 0$) with parallel homology. Now, we consider the antiparallel case and test both these results against MC simulations that exactly average the minimum energy described by Eqs. (4) and (10). Expressions of the average energy and the self-consistent equations that both $\Delta\bar{\Phi}_0$ and $\lambda_h$ satisfy are given in Appendix C of the Supplemental Material (and Ref. [43] for parallel homology). In Eq.(11), we have set the limits of integration to infinity, which is justified provided that $L_H / \lambda_h \gg 1$. One of the key results is that the approximation (Eq. (11)) predicts that, in the limit $L_H \to \infty$, the recognition energy per base pair approaches zero for the antiparallel case. This is indeed confirmed by the simulations, for which there is good agreement with the approximation (see Fig. 3 in the results section).

Unlike the rigid case, when $L_H / \lambda_h \gg 1$, there is effectively no difference in how we perform the order of averaging. Due to torsional flexibility, the correlation range $\lambda_h$ is finite, whereas for rigid molecules it is always infinite. The upshot is that over a length $|z - z'|$ that is larger than $\lambda_h$, the optimal phase difference $\Delta\Phi(z)$ (at $z$) loses memory of the pattern of base pair sequences at $z'$ described by $\delta\Omega_1(z') - \delta\Omega_2(z')$. This is due to the ability of the DNA fragments to torsionally adapt, at the cost of torsional elastic energy, so as to reduce their electrostatic energy. It does not matter where accumulation of phase mismatch occurs, as it only accumulates over length scales smaller than the correlation length $\lambda_h$, not over the whole fragment, unlike the rigid case.

## 3. Results

*Rigid Molecules*

In all calculations, we use the parameter values $a_1 = 0.015 k_B T / Å$ and $a_2 = 0.0035 k_B T / Å$ [50], previously used in [42]. We start by investigating the case where the torsional rigidity constant $C$ is infinitely large, as discussed in the first part of the theory section. In Fig.2 plots (at $x = 0$) of the recognition energies of antiparallel homologous and parallel homologous pairs are provided. Upon ensemble averaging over different sequences before minimizing the free energy, we find that the agreement between the analytical form of the recognition energies (see Eq. (8)) and those found from MC simulations is excellent, providing us with a useful test of the reliability of both. However, at present, only MC simulations can provide us with reliable quantitative results when we minimize the energy of a particular pairing of sequences before ensemble averaging over different sequence realizations. As discussed in the theory section, it is indeed possible to develop a perturbation expansion, for which we have calculated the leading order term (Eq. (9)). Unfortunately, there is little quantitative agreement between the approximation and simulation for the values of $L$ and $L_H$ considered here. The perturbation expansion, however, provides the correct qualitative trends. Therefore, we have only included this analytical result as part of the supplementary material.

The recognition energies between fragments containing parallel homologous segments and between antiparallel homologous segments are somewhat similar for homology lengths less than 250 base pairs. For larger homology lengths, e.g, at 2000 base pairs, the difference between these recognition energies becomes rather large. Indeed, upon examining Eq. (7) the large limit of increasing $L_H$, the recognition energy for parallel homology grows linearly with $L_H$. However, for the case of antiparallel homology (Eq. (8)), it approaches a constant value.

Upon minimizing the interaction energy with respect to pair orientation before ensemble averaging over different sequence realizations (different values of $\Delta\Phi_0$ for each), we find that the magnitude of the recognition energy for parallel homologous segments is somewhat reduced but increases slightly for the antiparallel case.

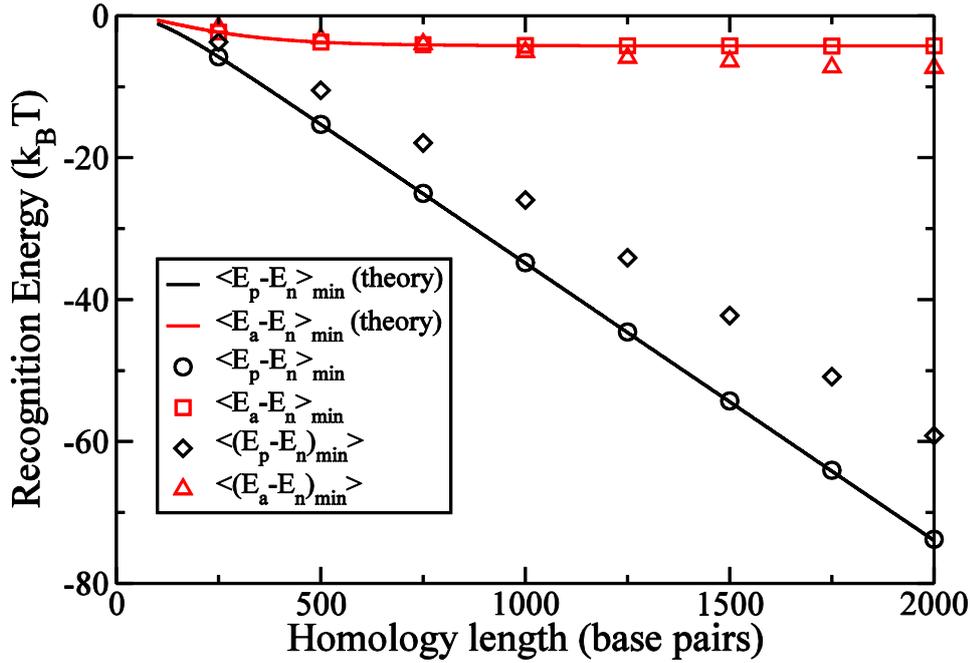

Figure 2. The recognition energy as a function of the homology length. The MC simulation data is shown as points, whereas theoretical curves for averaging before minimizing (from Eq. (7) and Eq.(8)) are presented by solid lines (all for $L_{NH}=0$). Here, in all cases, the red lines/points correspond to the energy difference between pairs containing segments in antiparallel alignment of homologous segments and completely uncorrelated sequences, while the black corresponds to the difference between pairs that contain sequences in parallel alignment and random pairs. For the simulation data, the squares and circles correspond to minimizing the energy after ensemble averaging over different sequences, while the diamonds and triangles correspond to minimizing the energy before sequence averaging. This difference between these is explained in the main text.

*Flexible molecules*

We now consider the molecules as torsionally flexible so that $C$ is finite. In the results that we present, we use a value of the torsional persistence length $C/k_bT = 725\text{Å}$ [51]. This choice is in good agreement with the largest DNA length considered in the simulations of Ref. [52].

In Figure 3 we plot the recognition energy for both the pairs containing parallel and antiparallel homology as a function of $x$, the shift between the two fragments making up a particular pair. We find a clear difference between the recognition energy wells for antiparallel homology pairing and those with parallel homology pairing. The well for the antiparallel case is much shallower and changes little with increasing homology length, whereas ,for the parallel pairing, it becomes much deeper with increasing homology length, $L_H$. We find, as in previous studies [33,43], that introducing torsional flexibility reduces the depth of the recognition well (to see this compare Figure 2 with Figure 3 at $x=0$). The is due to the torsional adaptation of the molecules, for which the energy for uncorrelated sequences now scales as $L_H$, as opposed to $\lambda_c^{(0)}$, which would be the case for long rigid fragments, hence reducing the pairing energy for completely random pairs. Consequently, since the fragments can adapt to each other, the difference between the total energy of homologous pairs and nonhomologous, uncorrelated pairs will be less pronounced. But, again, because such adaptation requires torsional elastic energy, the recognition energy, i.e., the difference in the energies between homologous and nonhomologous pairs, survives.

Furthermore, the width of the recognition well, for parallel homology, grows narrower as compared to that found for rigid molecules, originally considered in Ref. [42]. As discussed previously [43], one would expect that, for torsionally flexible molecules, the width of the well is effectively determined by the shorter correlation/adaptation length $\lambda_h$ (at $x = 0$), instead of the longer helical coherence length $\lambda_c^{(0)}$. As fragments slide along each other beyond the correlation length/adaptation length, one would expect that any information about the homology between them will effectively get washed out due to torsional adaptation of the molecules.

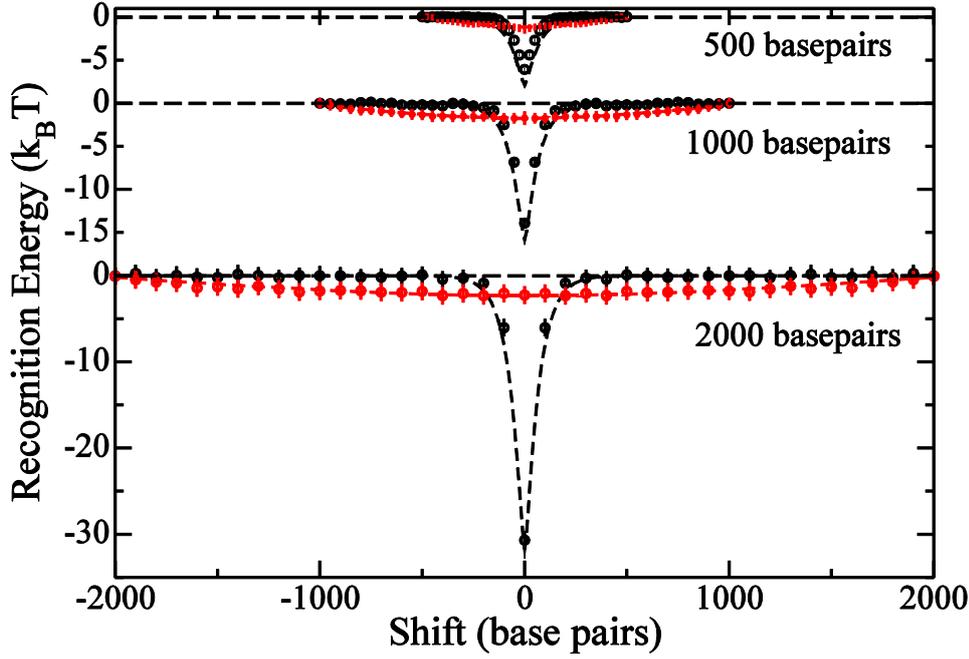

Figure 3. The recognition wells for various homology lengths (the value of $L_H$ is given in the top right corner of each panel, $L_{NH} = 0$) plotted as function of the shift, $x$ (offset for clarity's sake). The recognition energy for parallel homologous sequences is provided in black, whereas that for fragments containing antiparallel homologous sequences is provided in red. The points were found via MC simulations. For pairs containing antiparallel homology, the large $L_H$ theory predicts that the recognition energy is zero, shown as a flat dashed line. In the case for pairs with parallel homology the theoretical calculation of the recognition energy is given by the dashed line that passes through the simulation data, demonstrating good agreement with the MC simulation data. The MC data for antiparallel sequences, in red, were fit with a Lorentzian function to help guide the eye.

Next, we examine the recognition energy per base pair at $x = 0$ as a function of homology length $L_H$. Indeed, the theory [Eq. (11)] predicts that the recognition energy per base pair in the case of antiparallel homology should go to zero as $L_H \to \infty$, whereas for parallel homology it should tend to a constant value (see Appendix C of Supplemental Material). This indeed what is seen from the plots in Figure 4. A good fit to the MC numerical data of the recognition energy per base pair is given by the empirical formula $\varepsilon_{rec}(L_H) = (A + BL_H)^{-1}$.

We should add that in Appendix C of the supplemental material we included thermal fluctuations and have calculated the recognition well for the same parameter values, using also the values $L_H = 6800$ Å (corresponding to $2000$ base pairs) and $L_{NH} = 0$. We find that the recognition well seems to be affected very slightly when thermal fluctuations are included (see Appendix E of Supplemental Material).

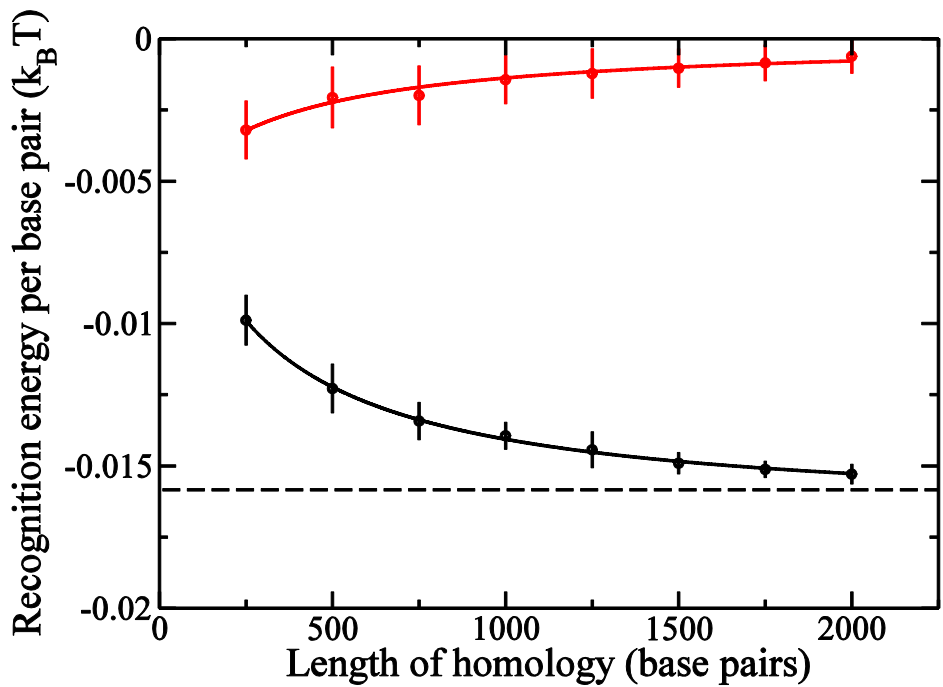

Figure 4. The recognition energy at the bottom of the well ($x = 0$, no shift) as a function of homology length. Here, the minimum value of the recognition well (at $x = 0$) is shown as a function of the length of homology $L_H$ for both the case of fragments containing parallel (black) and antiparallel (red) homologous segments (when nonhomologous segments flanking the homologous part are absent, $L_{NH} = 0$). The MC numerical data are fit very well by an empirical formula given in the text. The approximation for the recognition energy for the parallel homologous pairs, as described by Eq. (11), in the $L_H \to \infty$ limit is also provided in the plot (by the dashed black line). For antiparallel molecules this approximation predicts zero recognition energy per base pair with increasing homology length, conforming to the trend seen in the simulations.

## 4. Discussion

Intuitively, within the context of the recognition model, it was thought that DNA fragments containing homologous sequences in antiparallel alignment had the same average pairing energy as those with completely uncorrelated base pair tracts [42]. A core motivation for this study was realizing that this was not necessarily the case, strictly speaking, and that further analysis was needed. We have found that there is, indeed, a difference in pairing energy between antiparallel aligned homologous sequences and nonhomologous sequences, resulting in a small value of recognition energy (the difference between homologous and completely uncorrelated pairing energies). However, in the rigid case, for homology lengths fewer than 250 base pairs, the recognition energy at antiparallel orientation is of comparable magnitude to that of the parallel one [53]. The general trend for both rigid and flexible molecules is that the antiparallel recognition energy per base pair disappears with increasing homology length, while for the parallel one it tends towards a finite constant value.

In devising this study, we also sought to test the analytical approximations used in previous studies [42,43] against MC simulations. When optimizing the azimuthal orientations of the fragments after minimizing the energy for rigid molecules, we have found excellent agreement between the two. We have also investigated whether the order of averaging over different realizations of base pair sequence dependent distortions and optimizing the angle of rotation of the fragments about their long axis mattered. For rigid fragments the base pair helix disorder should be considered as quenched. This means that azimuthal orientations of the DNA fragments about their long axes should be optimized for each realization of the base pair dependent helix distortions first. The evaluation of the ensemble average of the energy should be considered next. However, swapping the order leads to simple analytical expressions, and we wanted to check the inaccuracy of this approximation. We have found a slight difference in recognition energy. Moreover, this difference vanished when considering long torsionally flexible molecules. In this case, we found excellent agreement between the analytical result and MC simulation. This suggests that the effect of changing the order of operations is, at most, significant only for homology lengths smaller than the adaptation length $\lambda_h$ (only DNA fragments smaller than $\lambda_h$ can be considered as being effectively rigid [33,54]). Although we have analytical approximations for both long and short homology lengths (rigid molecules), we have yet to build an approximation that interpolates between the two for the recognition well. This might be the subject of future work.

Through MC simulations, we are able to confirm that torsional flexibility reduced the depth of the recognition well as well as making it narrower. We also found that introducing thermal fluctuations (see Appendix D of the Supplemental Material) did little to change the shape of the well. Still the depth of the recognition well for long parallel homology fragments is much larger than $k_B T$ and the well is much wider than the Debye screening length in physiological solution (both as in previous works [42,43]), which one might naively expect from electrostatic interactions between DNA. If the recognition well was indeed too narrow or too shallow, one would expect that thermal fluctuations could disrupt an alignment of homologous genes due to the helix distortions. A very narrow recognition well might lead to a long initial homology search or missing the target, both impeding the whole process of homology recognition. On the other hand, if the recognition well was too wide, homologous genes may not reach ideal alignment, due to large thermal fluctuations about their optimal alignment; this could slow down the rate of homologous recombination and might reduce its fidelity. (But there might be the possibility of additional fine graining, discussed below).

Thus, our calculations suggest that natural sequence-related helix distortions could provide a reliable initial coarse-grained homology search by providing a large target for recognition, relying also on homology between sequences much larger than 10 base pairs. For sequences containing 10 base pairs or fewer our calculations suggest that this initial recognition step would be washed out by thermal fluctuations. This might prevent genes containing such small lengths of homology from efficiently recombining, except under specific conditions [55].

Global homology recognition through helix distortions may not be the complete search mechanism, even in the absence of strand breakage [56]. There is some evidence [46] to suggest the importance of an interaction that recognizes homology at the local level of individual base pairs, yet the nature of this interaction remains unclear. However, as homologous pairing was shown not to depend on the exact nature of the base pair text [46], those results seem to rule out the stem loop kissing mechanism [13,14,15,16]. The pairing energy from such a local base pair specific interaction, for double helices, would also be affected by helix distortions, as the base pairs would have to align in an optimal orientation for this type of pairing interaction to be sufficiently strong. The theoretical implications of such a contribution to homology recognition have yet to be investigated, and may be the topic for further work [56]. However, the contribution of such an interaction alone would be likely to lead to a very narrow recognition well, which would be determined by short range interactions. This mechanism alone would likely lead to a long homology search [56]. Thus, we think that the recognition mechanism due to base-pair specific helix distortions is important in providing a fast initial recognition step.

On the experimental side, one of us is currently investigating differences between supercoiled DNA *plasmids* that contain homologous segments, in either parallel or antiparallel alignment, and plasmids in which the sequence texts are completely uncorrelated. Indeed, if interaction forces that depend on DNA helix structure are indeed important, on the basis of the calculations presented here, we would expect a significant difference in conformation and dynamic properties (assessed by different experimental methods) between plasmids with parallel homology and plasmids with completely uncorrelated sequences, but not between plasmids containing antiparallel homologous segments and completely uncorrelated ones. In developing more accurate models to describe the difference between such plasmids, we hope to take account of the braiding of DNA through closed supercoiling. Currently, a statistical mechanical model for closed loop supercoiling, including interactions that depend on helix structure, is being developed [57]. This model is based on the previous work of Refs. [35,49,58,59,60]. We hope to use this model to build upon this investigation, further rationalizing the experiments underway. There are also further experiments that can be performed. The experiments of Ref. [46] were performed *in vivo*, although in the absence of the ME13 protein, an analogue of RecA specific to the species considered there (suggesting that single strand breakage was not important in the repeat point mutation process under study). Thus, future experimental work, *in vitro* in a protein free environment, should also consider building DNA pairs that contain special periodic arrays of sequence homology similar to those of Ref. [46]. Such specially designed experiments, in controlled conditions, may help to further elucidate the relative importance of various homology pairing mechanisms [56].

## 5. Concluding points

In agreement with predictions, we have shown that there is little difference between the interaction energy of a pair of nonhomologous sequences and a pair of antiparallel homologous ones. This fact is important in designing single molecule experiments. But in addition: our analysis suggests that the homology recognition mechanism described herein significantly favours the correct, parallel alignment of homologous DNA.

We have discovered that a slight difference exists in the results for the calculated recognition energy for rigid molecules, depending on whether we optimized the azimuthal orientation of the molecules for each sequence dependent pattern of helix distortions and then performed an ensemble average of the energy as compared to averaging the energy before optimizing the azimuthal orientation. This difference, however, vanishes for long torsionally flexible molecules.

The MC simulations performed in this work confirm our previous results accounting for the estimated torsional flexibility of DNA, namely that the recognition well grows shallower and narrower. This, however, does not preclude its biological significance.

## Acknowledgements

The authors would like to thank Sergey Leikin (NICHHD, NIH Bethesda), Takashi Ohyama (Waseda University), Mara Prentiss (Harvard University), and Kenichi Yoshikawa (Doshisha University) for useful discussions. DJL and AAK would like to acknowledge the support of the HFSP grant "Chiral effects in DNA supercoiling"(grant RGP0049/2010-C102).

# Supplemental Material

## Appendix A. Averaging before minimizing

Here we consider performing the ensemble average over sequence realizations before optimizing the interaction energy over $\Delta\bar{\Phi}$. Therefore, our starting point is Eqs. (5) and (6) of the main text. Here, we generalize to consider $L_{NH} \neq 0$, but for simplicity we only study the $x = 0$ case, for which there is complete overlap of the homologous sections of the interacting DNA molecules. We will consider the three types of paired fragments in turn. We start first with the two sets of base pairs completely uncorrelated, followed by the fragments containing parallel homology, and finish with fragments containing antiparallel homology. Upon calculating their respective energies, we will obtain the expressions for the recognition energy, the difference in the energies between the homologous and completely non-homologous (uncorrelated) cases.

*Uncorrelated case*

We combine Eqs. (5) and (6) of the main text so that we have

$$\Delta\Phi(z) = \Delta\bar{\Phi}_0 + \Delta\tilde{\Phi}(z), \tag{A.1}$$

where

$$\Delta\tilde{\Phi}(z) = \frac{1}{h}\int_d^z \left[\delta\Omega_1(z') - \delta\Omega_2(z')\right]dz', \qquad d < z < L, \tag{A.2}$$

$$\Delta\tilde{\Phi}(z) = -\frac{1}{h}\int_z^d \left[\delta\Omega_1(z') - \delta\Omega_2(z')\right]dz', \qquad 0 < z < d. \tag{A.3}$$

Here, $d$ is the location where $\Delta\Phi(d) = \Delta\bar{\Phi}_0$. From Eq. (4) of the main text, the average interaction energy is given by the expression

$$\langle E_{\text{int}}\rangle_{\delta\Omega} = La_0(R) + \int_0^L dz \left(-a_1(R)\langle\cos\Delta\Phi(z)\rangle_{\delta\Omega} + a_2(R)\langle\cos 2\Delta\Phi(z)\rangle_{\delta\Omega}\right). \tag{A.4}$$

The angular bracket refers to averaging over all the possible realizations of both $\delta\Omega_1(z')$ and $\delta\Omega_2(z')$, the sequence-dependent deviations of the base-pair twist away from the average value, which describes an ideal helix. Since Eqs. (A.2) and (A.3) are linear in $\delta\Omega_\mu(z)$, with $\delta\Omega_\mu(z)$ Gaussian distributed, the cosine averages can be written as

$$\langle\cos n\Delta\Phi(z)\rangle_{\delta\Omega} = \cos n\Delta\bar{\Phi}_0 \exp\left(-\frac{n^2}{2}\langle\Delta\tilde{\Phi}(z)^2\rangle_{\delta\Omega}\right). \tag{A.5}$$

Using Eqs. (A.2) and (A.3), we can write

$$\langle\Delta\tilde{\Phi}(z)^2\rangle_{\delta\Omega} = \frac{1}{h^2}\int_d^z dz'\int_d^z dz'' \langle[\delta\Omega_1(z')-\delta\Omega_2(z')][\delta\Omega_1(z'')-\delta\Omega_2(z'')]\rangle_{\delta\Omega}, \qquad d<z<L, \tag{A.6}$$

$$\langle\Delta\tilde{\Phi}(z)^2\rangle_{\delta\Omega} = \frac{1}{h^2}\int_z^d dz'\int_z^d dz'' \langle[\delta\Omega_1(z')-\delta\Omega_2(z')][\delta\Omega_1(z'')-\delta\Omega_2(z'')]\rangle_{\delta\Omega}, \qquad 0<z<d. \tag{A.7}$$

For the uncorrelated case, for all $z$, $\langle\delta\Omega_1(z'')\delta\Omega_2(z')\rangle_{\delta\Omega} = 0$. Combining this relationship with Eq. (3) of the main text yields

$$\langle\Delta\tilde{\Phi}(z)^2\rangle_{\delta\Omega} = \frac{2}{\lambda_c^{(0)}}\int_d^z dz'\int_d^z dz''\delta(z'-z'') = \frac{2(z-d)}{\lambda_c^{(0)}}, \qquad d<z<L, \tag{A.8}$$

$$\langle\Delta\tilde{\Phi}(z)^2\rangle_{\delta\Omega} = \frac{2}{\lambda_c^{(0)}}\int_z^d dz'\int_z^d dz''\delta(z'-z'') = \frac{2(d-z)}{\lambda_c^{(0)}}, \qquad 0<z<d. \tag{A.9}$$

The average energy interaction energy (A.5) can thus be evaluated as

$$\langle E_{\text{int}}\rangle_{\delta\Omega} = La_0(R) - \lambda_c^{(0)} a_1(R)\cos\Delta\bar{\Phi}_0 \left[2 - \exp\left(-\frac{d}{\lambda_c^{(0)}}\right) - \exp\left(-\frac{L-d}{\lambda_c^{(0)}}\right)\right]$$
$$+ \frac{\lambda_c^{(0)} a_2(R)}{4}\cos 2\Delta\bar{\Phi}_0 \left[2 - \exp\left(-\frac{4d}{\lambda_c^{(0)}}\right) - \exp\left(-\frac{4(L-d)}{\lambda_c^{(0)}}\right)\right]. \tag{A.10}$$

Minimization of Eq. (A.10) with respect to $d$ yields

$$0 = -a_1(R)\cos\Delta\bar{\Phi}_0\left[\exp\left(-\frac{d}{\lambda_c^{(0)}}\right) - \exp\left(-\frac{L-d}{\lambda_c^{(0)}}\right)\right]$$
$$+ a_2(R)\cos 2\Delta\bar{\Phi}_0\left[\exp\left(-\frac{4d}{\lambda_c^{(0)}}\right) - \exp\left(-\frac{4(L-d)}{\lambda_c^{(0)}}\right)\right]. \tag{A.11}$$

This provides the solution for the location, $d = L/2$, of the phase alignment that minimizes the average energy. Also, minimization of Eq. (A.10) with respect to $\Delta\bar{\Phi}_0$ yields

$$\cos\Delta\bar{\Phi}_0 = \frac{a_1(R)}{4a_2(R)} \frac{\left[1-\exp\left(-\frac{L}{2\lambda_c^{(0)}}\right)\right]}{\left[1-\exp\left(-\frac{2L}{\lambda_c^{(0)}}\right)\right]} \leq 1, \quad \text{or} \quad \cos\Delta\bar{\Phi}_0 = 1. \quad (A.12)$$

For the values of the *a* coefficients used in our simulations, we need only consider the latter solution, $\cos\Delta\bar{\Phi}_0 = 1$, so

$$\langle E_{\text{int}} \rangle_{\delta\Omega} = La_0(R) - 2\lambda_c^{(0)} a_1(R)\left[1-\exp\left(-\frac{L}{2\lambda_c^{(0)}}\right)\right] + \frac{\lambda_c^{(0)} a_2(R)}{2}\left[1-\exp\left(-\frac{2L}{\lambda_c^{(0)}}\right)\right]. \quad (A.13)$$

*Parallel Homologous case*

For the parallel homologous case, Eqs. (A.4)-(A.7) can still be utilized, however the next step to determine the interaction energy requires a different formulation. Crucially, the deviation of the twist angles are now the same for the two fragments, $\delta\Omega_1(z') = \delta\Omega_2(z')$, when $L_{NH} < z' < L_H + L_{NH}$, but as with the random case, $\langle\delta\Omega_1(z')\delta\Omega_2(z'')\rangle = 0$ elsewhere. In what follows, we will assume that the optimum location for the twist angle alignment occurs again at $d = L/2$ on symmetry grounds alone; rigorously proving this is complicated so we refrain from presenting this here.

Equations (A.6) and (A.7), for parallel homologous fragments, become

$$\langle\Delta\tilde{\Phi}(z)^2\rangle_{\delta\Omega} = \frac{2(z-L_H-L_{NH})}{\lambda_c}, \quad \text{when} \quad z > L_H + L_{NH}, \quad (A.14)$$

$$\langle\Delta\tilde{\Phi}(z)^2\rangle_{\delta\Omega} = 0, \quad \text{when} \quad L_H + L_{NH} > z > L_{NH}, \quad (A.15)$$

$$\langle\Delta\tilde{\Phi}(z)^2\rangle_{\delta\Omega} = \frac{2(L_{NH}-z)}{\lambda_c}, \quad \text{when} \quad z < L_{NH}. \quad (A.16)$$

Thus, for the average energy, we obtain from Eqs. (A.4), (A.5), (A.14)-(A.16)

$$\langle E_{\text{int}}\rangle_{\delta\Omega} = La_0(R) - L_H\left(a_1(R)\cos\Delta\bar{\Phi}_0 - a_2(R)\cos 2\Delta\bar{\Phi}_0\right)$$
$$-2\lambda_c\left[a_1(R)\cos\Delta\bar{\Phi}_0\left(1-\exp\left(-\frac{L_{NH}}{\lambda_c}\right)\right) - \frac{a_2(R)}{4}\cos 2\Delta\bar{\Phi}_0\left(1-\exp\left(-\frac{4L_{NH}}{\lambda_c}\right)\right)\right]. \quad (A.17)$$

Setting $\cos \Delta\bar{\Phi}_0 = 1$ in Eq. (A.17) yields

$$\langle E_{int} \rangle_{\partial\Omega} = La_0(R) - L_H(a_1(R) - a_2(R)) - 2\lambda_c \left[ a_1(R)\left(1 - \exp\left(-\frac{L_{NH}}{\lambda_c}\right)\right) - \frac{a_2(R)}{4}\left(1 - \exp\left(-\frac{4L_{NH}}{\lambda_c}\right)\right) \right]. \quad (A.18)$$

*Antiparallel Homologous case*

For the antiparallel homologous case, again Eqs. (A.4)-(A.7) hold. However, now we have $\delta\Omega_1(z') = \delta\Omega_2(2L_{NH} + L_H - z')$ when $L_{NH} < z' < L_H + L_{NH}$ and $\langle \delta\Omega_1(z')\delta\Omega_2(z'') \rangle = 0$ elsewhere.

Upon on evaluation of Eqs. (A.6) and (A.7) we find new expressions for $\langle \Delta\tilde{\Phi}(z)^2 \rangle_{\partial\Omega}$. For $0 > d > L_{NH}$ we obtain

$$\langle \Delta\tilde{\Phi}(z)^2 \rangle_{\partial\Omega} = \frac{2}{\lambda_c^{(0)}} |d - z|, \qquad 0 < z < L_{NH} + L_H/2, \quad (A.19)$$

$$\langle \Delta\tilde{\Phi}(z)^2 \rangle_{\partial\Omega} = \frac{2}{\lambda_c^{(0)}} (L_H + 2L_{NH} - d - z), \qquad L_{NH} + L_H/2 < z < L_{NH} + L_H, \quad (A.20)$$

$$\langle \Delta\tilde{\Phi}(z)^2 \rangle_{\partial\Omega} = \frac{2}{\lambda_c^{(0)}} (z - d - L_H), \qquad L_{NH} + L_H < z < 2L_{NH} + L_H, \quad (A.21)$$

When $L_{NH} < d < L_{NH} + \frac{L_H}{2}$,

$$\langle \Delta\tilde{\Phi}(z)^2 \rangle_{\partial\Omega} = \frac{2}{\lambda_c^{(0)}} |d - z|, \qquad 0 < z < L_{NH} + L_H/2, \quad (A.22)$$

$$\langle \Delta\tilde{\Phi}(z)^2 \rangle_{\partial\Omega} = \frac{2}{\lambda_c^{(0)}} |L_H + 2L_{NH} - d - z|, \qquad L_{NH} + L_H/2 < z < 2L_{NH} + L_H. \quad (A.23)$$

When $L_{NH} + L_H > d > L_{NH} + \frac{L_H}{2}$,

$$\langle \Delta\tilde{\Phi}(z)^2 \rangle_{\partial\Omega} = \frac{2}{\lambda_c^{(0)}} |z - d|, \qquad L_{NH} + L_H/2 < z < 2L_{NH} + L_H, \quad (A.24)$$

$$\langle \Delta\tilde{\Phi}(z)^2 \rangle_{\partial\Omega} = \frac{2}{\lambda_c^{(0)}} |z + d - L_H - 2L_{NH}|, \qquad 0 < z < L_{NH} + L_{NH}/2. \quad (A.25)$$

Finally, when $2L_{NH} + L_H > d > L_{NH} + L_H$

$$\langle \Delta\tilde{\Phi}(z)^2 \rangle_{\partial\Omega} = \frac{2}{\lambda_c^{(0)}} |z - d|, \qquad L_{NH} + L_H/2 < z < 2L_{NH} + L_H, \quad (A.26)$$

$$\left\langle \Delta\tilde{\Phi}(z)^2 \right\rangle_{\partial\Omega} = \frac{2}{\lambda_c^{(0)}}(z+d-2L_{NH}-L_H), \qquad L_{NH} < z < L_{NH} + L_H/2, \qquad (A.27)$$

$$\left\langle \Delta\tilde{\Phi}(z)^2 \right\rangle_{\partial\Omega} = \frac{2}{\lambda_c^{(0)}}(d-z-L_H), \qquad 0 < z < L_{NH}. \qquad (A.28)$$

To evaluate for the ensemble averaged energy (Eq. (A.5)) for this case, we must evaluate the integral $X_n$, defined as

$$X_n = \int_0^L dz\, \exp\left(-\frac{n^2 \left\langle \Delta\tilde{\Phi}(z)^2 \right\rangle_{\partial\Omega}}{2}\right). \qquad (A.29)$$

We first evaluate $X_n$ for the case where $0 < d < L_{NH}$. Substituting Eqs. (A.19)-(A.21) into Eq. (A.29) yields

$$X_n = \frac{\lambda_c^{(0)}}{n^2}\left[2-\exp\left(-\frac{n^2 d}{\lambda_c^{(0)}}\right) - 2\exp\left(\frac{n^2(d-L_{NH}-L_H/2)}{\lambda_c^{(0)}}\right) + 2\exp\left(\frac{n^2(d-L_{NH})}{\lambda_c^{(0)}}\right) - \exp\left(\frac{n^2(d-2L_{NH})}{\lambda_c^{(0)}}\right)\right]. \qquad (A.30)$$

Next, we consider the case $L_{NH} < d < L_{NH} + L_H/2$. Here, upon substituting Eqs. (A.22) and (A.25) into Eq. (A.29), we obtain

$$X_n = \frac{\lambda_c^{(0)}}{n^2}\left[4 - 2\exp\left(-\frac{n^2 d}{\lambda_c^{(0)}}\right) - 2\exp\left(\frac{n^2(d-L_{NH}-L_H/2)}{\lambda_c^{(0)}}\right)\right]. \qquad (A.31)$$

In the case where $L_{NH} + L_H/2 < d < L_{NH} + L_H$, upon substituting Eqs. (A.26) and (A.25) into Eq. (A.29), we find

$$X_n = \frac{\lambda_c^{(0)}}{n^2}\left[4 - 2\exp\left(-\frac{(L_H+2L_{NH}-d)n^2}{\lambda_c^{(0)}}\right) - 2\exp\left(-\frac{n^2(d-L_{NH}-L_H/2)}{\lambda_c^{(0)}}\right)\right]. \qquad (A.32)$$

Last of all, we obtain for $L_{NH} + L_H < d < 2L_{NH} + L_H$, by substituting Eqs. (A.26)-(A.28) into Eq. (A.29), the expression

$$X_n = \frac{\lambda_c^{(0)}}{n^2}\left[2 - \exp\left(-\frac{n^2(L_H+2L_{NH}-d)}{\lambda_c^{(0)}}\right) - 2\exp\left(-\frac{n^2(d-L_{NH}-L_H/2)}{\lambda_c^{(0)}}\right) \right.$$
$$\left. + 2\exp\left(-\frac{n^2(d-L_{NH}-L_H)}{\lambda_c^{(0)}}\right) - \exp\left(-\frac{n^2(d-L_H)}{\lambda_c^{(0)}}\right)\right]. \qquad (A.33)$$

We will assume that average energy is minimized when $X_n$ is maximized (which should be true when $a_1(R) > 4a_2(R)$, the case that we consider) since Eqs. (A.30) and (A.33), valid for $0 < d < L_{NH}$ and $L_{NH} + L_H < d < 2L_{NH} + L_H$ respectively, possess no turning points (no points where the derivative respect to $d$ is zero) in these ranges. Indeed, Eq. (A.30) monotonically increases as a function of $d$, as $d \to L_{NH}$, while Eq. (A.33) monotonically decreases as $d \to L_{NH} + L_H$. Turning points only occur for the expressions valid for $L_{NH} < d < L_{NH} + L_H/2$ and $L_{NH} + L_H/2 < d < L_{NH} + L_H$, Eqs.(A.31) and (A.32). The turning points (extrema) thus satisfy the equations

$$0 = \frac{dX_n}{d(d)} = 2\exp\left(-\frac{dn^2}{\lambda_c^{(0)}}\right) - 2\exp\left(\frac{n^2(d - L_{NH} - L_H/2)}{\lambda_c^{(0)}}\right), \quad (A.34)$$

and

$$0 = \frac{dX_n}{d(d)} = 2\exp\left(-\frac{(L_H + 2L_{NH} - d)n^2}{\lambda_c^{(0)}}\right) - 2\exp\left(-\frac{n^2(d - L_{NH} - L_H/2)}{\lambda_c^{(0)}}\right). \quad (A.35)$$

The solutions to Eq. (A.34) and (A.35) are $d = L_{NH}/2 + L_H/4$ and $d = 3L_{NH}/2 + 3L_H/4$. Therefore, provided that $L_H > 2L_{NH}$ these two values of $d$ maximize $X_n$. When $L_H < 2L_{NH}$ there are no turning points in the complete piecewise function $X_n$, instead $X_n$ is maximized where there are discontinuities in the derivative that lie at $d = L_{NH}$ and $d = L_{NH} + L_H$. Thus, for $a_1(R) > 4a_2(R)$, the minimized energy is

$$\langle E_{int}\rangle_{\partial\Omega} = a_0(R)(2L_{NH} + L_H) + \left\{-4a_1(R)\lambda_c^{(0)}\left[1-\exp\left(-\frac{L_{NH} + L_H/2}{2\lambda_c^{(0)}}\right)\right]\right.$$
$$+a_2(R)\lambda_c^{(0)}\left[1-\exp\left(-\frac{(2L_{NH} + L_H)}{\lambda_c^{(0)}}\right)\right]\right\}\theta(L_H - 2L_{NH})$$
$$\left\{-a_1(R)\lambda_c^{(0)}\left[4 - 2\exp\left(-\frac{L_{NH}}{\lambda_c^{(0)}}\right) - 2\exp\left(-\frac{L_H}{2\lambda_c^{(0)}}\right)\right]\right.$$
$$+a_2(R)\lambda_c^{(0)}\left[1 - \frac{1}{2}\exp\left(-\frac{4L_{NH}}{\lambda_c^{(0)}}\right) - \frac{1}{2}\exp\left(-\frac{2L_H}{\lambda_c^{(0)}}\right)\right]\right\}\theta(2L_{NH} - L_H). \quad (A.36)$$

*Recognition Energies*

We now obtain the recognition energies by taking the difference between the interaction energies of molecules with parallel homology and those with uncorrelated sequences. Subtracting Eq. (A.13) from (A.18) yields

$$E_{rec} = -L_H(a_1(R) - a_2(R))$$
$$+2\lambda_c\left[a_1(R)\exp\left(-\frac{L_{NH}}{\lambda_c}\right)\left(1 - \exp\left(-\frac{L_H}{2\lambda_c}\right)\right) - \frac{a_2(R)}{4}\exp\left(-\frac{4L_{NH}}{\lambda_c}\right)\left(1 - \exp\left(-\frac{2L_H}{\lambda_c}\right)\right)\right]. \quad (A.37)$$

The recognition energy for molecules with antiparallel homologous alignment is found similarly, by subtracting Eq. (A.13) from (A.36):

$$E_{rec} = \left\{ -a_1(R)\lambda_c^{(0)} \left[ 2 - 4\exp\left(-\frac{L_{NH} + L_H/2}{2\lambda_c^{(0)}}\right) + 2\exp\left(-\frac{L_{NH} + L_H/2}{\lambda_c^{(0)}}\right) \right] \right.$$

$$\left. + a_2(R)\lambda_c^{(0)} \left[ \frac{1}{2} - \exp\left(-\frac{2L_{NH} + L_H}{\lambda_c^{(0)}}\right) + \frac{1}{2}\exp\left(-\frac{4L_{NH} + 2L_H}{\lambda_c^{(0)}}\right) \right] \right\} \theta(L_H - 2L_{NH}) +$$

$$\left\{ -a_1(R)\lambda_c^{(0)} \left[ 2 - 2\exp\left(-\frac{L_{NH}}{\lambda_c^{(0)}}\right) - 2\exp\left(-\frac{L_H}{2\lambda_c^{(0)}}\right) + 2\exp\left(-\frac{L_{NH} + L_H/2}{\lambda_c^{(0)}}\right) \right] \right.$$

$$\left. + a_2(R)\lambda_c^{(0)} \left[ \frac{1}{2} - \frac{1}{2}\exp\left(-\frac{4L_{NH}}{\lambda_c^{(0)}}\right) - \frac{1}{2}\exp\left(-\frac{2L_H}{\lambda_c^{(0)}}\right) + \frac{1}{2}\exp\left(-\frac{4L_{NH} + 2L_H}{\lambda_c^{(0)}}\right) \right] \right\} \theta(2L_{NH} - L_H).$$

(A.38)

Upon considering only the interaction energy over the region of homologous overlap, we set $L_{NH} = 0$, giving Eqs. (7) and (8) of the main text.

## Appendix B. Minimizing before averaging

### *General Considerations*

Now, let us consider minimizing the energy to find the optimal azimuthal orientation of each type of pair, $\Delta\Phi_0$ for each realization of $\delta\Omega_1(z)$ and $\delta\Omega_2(z)$ before ensemble averaging over different sequence twist angle realizations. Minimizing the unaveraged interaction energy (Eq. (4) of the main text) with respect to $\Delta\Phi_0$ yields the following integral equation

$$0 = \int_0^L dz \left( -a_1(R)\left[ \sin(\Delta\tilde{\Phi}(z))\cos(\Delta\Phi_0) + \cos(\Delta\tilde{\Phi}(z))\sin(\Delta\Phi_0) \right] \right)$$
$$+ \int_0^L dz \left( a_2(R)\left[ \sin(2\Delta\tilde{\Phi}(z))\cos(2\Delta\Phi_0) + \cos(2\Delta\tilde{\Phi}(z))\sin(2\Delta\Phi_0) \right] \right),$$
(B.1)

where

$$\Delta\tilde{\Phi}(z) = \frac{1}{h}\int_0^z dz' \left( \delta\Omega_1(z') - \delta\Omega_2(z') \right).$$
(B.2)

In principle, one can solve Eq. (B.1) for $\cos(\Delta\Phi_0)$ terms of $\int_0^L dz \sin(\Delta\tilde{\Phi}(z))$ by solving a quadratic equation. However, such a solution would be cumbersome, and it is probably not possible to obtain an analytical expression of the average of the energy over different twist angle realizations $\delta\Omega_\mu(z')$. In what follows, we carry out an analysis by simply treating $\Delta\tilde{\Phi}(z)$ as a small perturbation from its average value [61]. In other words, we write $\Delta\Phi_0 = \Delta\bar{\Phi}_0 + \chi$ where now $\Delta\bar{\Phi}_0$ is the value of that minimizes the interaction energy when $(\delta\Omega_1(z') - \delta\Omega_2(z')) = 0$, which is given by

$$\cos\Delta\bar{\Phi}_0 = \frac{a_1(R)}{4a_2(R)} \leq 1, \qquad \text{or} \qquad \cos\Delta\bar{\Phi}_0 = 1. \tag{B.3}$$

Here, $\chi$ is the correction to $\Delta\Phi_0$ due to $\Delta\tilde{\Phi}(z)$. We also expand out in powers of $\chi$. From Eq. (B.1), this expansion yields, to leading order,

$$0 = -a_1(R)\sin(\Delta\bar{\Phi}_0) + 2a_2(R)\sin(2\Delta\bar{\Phi}_0) + \int_0^L dz \left(-a_1(R)\left(\frac{1}{h}\int_0^z dz'(\delta\Omega_1(z') - \delta\Omega_2(z'))\right) + \chi\right)\cos(\Delta\bar{\Phi}_0)$$

$$+ \int_0^L dz \left(2a_2(R)\left(\frac{1}{h}\int_0^z dz'(\delta\Omega_1(z') - \delta\Omega_2(z'))\right) + 2\chi\right)\cos(2\Delta\bar{\Phi}_0).$$

(B.4)

Since we require that $-a_1(R)\sin(\Delta\bar{\Phi}_0) + 2a_2(R)\sin(2\Delta\bar{\Phi}_0) = 0$, we obtain from Eq. (B.4)

$$\chi = -\frac{1}{hL}\int_0^L dz \int_0^z dz'(\delta\Omega_1(z') - \delta\Omega_2(z')). \tag{B.5}$$

Thus we obtain the general result Eq. (9) of the main text. Again, let us deal with each of the three types of fragment pairs in turn, starting again with completely uncorrelated random sequences.

### *Random Sequences*

Both Eqs. (A.4) and (A.5) still apply. Therefore, we again start by computing the average value $\langle\Delta\tilde{\Phi}(z)^2\rangle_{\delta\Omega}$. We can first write this as

$$\langle\Delta\tilde{\Phi}(z)^2\rangle_{\delta\Omega} = \frac{1}{h^2L^2}\int_0^L dz \int_0^L dz' \int_0^z ds \int_0^{z'} ds' \langle(\delta\Omega_1(s) - \delta\Omega_2(s))(\delta\Omega_1(s') - \delta\Omega_2(s'))\rangle_{\delta\Omega}$$

$$-\frac{2}{h^2L}\int_0^L dz' \int_0^z ds \int_0^{z'} ds' \langle(\delta\Omega_1(s) - \delta\Omega_2(s))(\delta\Omega_1(s') - \delta\Omega_2(s'))\rangle_{\delta\Omega} \tag{B.6}$$

$$+\frac{1}{h^2}\int_0^z ds \int_0^z ds' \langle(\delta\Omega_1(s) - \delta\Omega_2(s))(\delta\Omega_1(s') - \delta\Omega_2(s'))\rangle_{\delta\Omega}.$$

Now we, again, use the fact that for completely uncorrelated sequences $\langle\delta\Omega_1(z')\delta\Omega_2(z)\rangle = 0$. Therefore, using Eq. (3) of the main text, we obtain a positive definite quadratic form,

$$\langle \Delta\tilde{\Phi}(z)^2 \rangle_{\delta\Omega} = \frac{2L}{3\lambda_c^{(0)}} + \frac{2z^2}{L\lambda_c^{(0)}} - \frac{2}{\lambda_c^{(0)}} z. \tag{B.7}$$

Let us consider again the integral $X_n$ as defined by Eqs. (A.29). Using Eq. (B.7), this can be written as

$$X_n = \exp\left(-\frac{Ln^2}{3\lambda_c^{(0)}}\right)\int_0^L \exp\left(-\frac{n^2 z^2}{L\lambda_c^{(0)}} + \frac{n^2 z}{\lambda_c^{(0)}}\right) dz = \exp\left(-\frac{Ln^2}{12\lambda_c^{(0)}}\right)\sqrt{\frac{\pi L \lambda_c^{(0)}}{n^2}}\, \mathrm{erf}\left(\sqrt{\frac{Ln^2}{4\lambda_c^{(0)}}}\right) \equiv X_r\left(L, \frac{\lambda_c^{(0)}}{n^2}\right). \tag{B.8}$$

Thus, we can write the average energy (given by Eq. (A.4)) as

$$\langle E_{\mathrm{int}} \rangle_\Omega = La_0 - La_1 X_r\left(L, \lambda_c^{(0)}\right)\cos\Delta\bar{\Phi}_0 + La_2 X_r\left(L, \lambda_c^{(0)}/4\right)\cos 2\Delta\bar{\Phi}_0. \tag{B.9}$$

### *Imbedded Parallel Homologous Sequences*

Here, as before in Appendix A, $\delta\Omega_1(z) = \delta\Omega_2(z)$ for $L_{NH} < z < L_H + L_{NH}$, otherwise $\langle\delta\Omega_1(z)\delta\Omega_2(z')\rangle_{\delta\Omega} = 0$. We first consider the average $\langle\chi^2\rangle_{\delta\Omega}$, which we can write, using Eq. (B.5), as

$$\langle\chi^2\rangle_{\delta\Omega} = \frac{1}{h^2(L_H + 2L_{NH})^2}\left[(L_H + L_{NH})^2 \int_0^{L_{NH}} ds \int_0^{L_{NH}} ds'\, A(s,s') + 2(L_H + L_{NH})\int_0^{L_{NH}} dz \int_0^{L_{NH}} ds \int_0^z ds'\, A(s,s')\right.$$

$$\left. + \int_0^{L_{NH}} dz' \int_0^{L_{NH}} dz \int_0^z ds \int_0^{z'} ds'\, A(s,s') + \int_{L_H+L_{NH}}^{2L_{NH}+L_H} dz' \int_{L_H+L_{NH}}^{2L_{NH}+L_H} dz \int_{L_H+L_{NH}}^z ds \int_{L_H+L_{NH}}^{z'} ds'\, A(s,s')\right], \tag{B.10}$$

where

$$A(s,s') = \langle\delta\Omega_1(s)\delta\Omega_1(s')\rangle_{\delta\Omega} + \langle\delta\Omega_2(s)\delta\Omega_2(s')\rangle_{\delta\Omega}. \tag{B.11}$$

Using Eq. (3) of the main text we may evaluate Eqs. (B.10). This yields

$$\langle\chi^2\rangle_{\delta\Omega} = \frac{2L_{NH}(L_H + L_{NH})}{\lambda_c^{(0)}(L_H + 2L_{NH})} + \frac{4L_{NH}^3}{3\lambda_c^{(0)}(L_H + 2L_{NH})^2}. \tag{B.12}$$

Next, let us again evaluate $\langle\Delta\tilde{\Phi}(z)^2\rangle$. For $0 < z < L_{NH}$,

$$\langle\Delta\tilde{\Phi}(z)^2\rangle_{\delta\Omega} = \frac{1}{h^2}\int_0^s ds \int_0^s ds'\, A(s,s') - \frac{2}{(L_H + 2L_{NH})h^2}\left[\int_0^{L_{NH}} dz' \int_0^z ds \int_0^{z'} ds'\, A(s,s')\right.$$

$$\left. -(L_H + L_{NH})\int_0^z ds \int_0^{L_{NH}} ds'\, A(s,s') - \int_{L_{NH}+L_H}^{2L_{NH}+L_H} dz' \int_0^z ds \int_0^{z'} ds'\, A(s,s')\right] + \langle\chi^2\rangle_{\delta\Omega}, \tag{B.13}$$

and for $L_{NH} < z < L_{NH} + L_H$.

$$\langle \Delta\tilde{\Phi}(z)^2 \rangle_{\partial\Omega} = \frac{1}{h^2} \int_0^{L_{NH}} ds \int_0^{L_{NH}} ds' A(s,s') - \frac{2}{(L_H + 2L_{NH})h^2} \left[ \int_0^{L_{NH}} dz' \int_0^{L_{NH}} ds \int_0^{z'} ds' A(s,s') \right.$$

$$\left. -(L_H + L_{NH}) \int_0^{L_{NH}} ds \int_0^{L_{NH}} ds' A(s,s') - \int_{L_{NH}+L_H}^{2L_{NH}+L_H} dz' \int_0^{L_{NH}} ds \int_0^{z'} ds' A(s,s') \right] + \langle \chi^2 \rangle_{\partial\Omega},$$

(B.14)

and finally for $L_{NH} + L_H < z < 2L_{NH} + L_H$,

$$\langle \Delta\tilde{\Phi}(z)^2 \rangle_{\partial\Omega} = \frac{1}{h^2} \left[ \int_0^{L_{NH}} ds \int_0^{L_{NH}} ds' A(s,s') + \int_{L_{NH}+L}^{z} ds \int_{L_{NH}+L_H}^{z} ds' A(s,s') \right]$$

$$- \frac{2}{(L_H + 2L_{NH})h^2} \left[ \int_0^{L_{NH}} dz' \int_{L_{NH}+L_H}^{z} ds \int_0^{z'} ds' A(s,s') + (L_H + L_{NH}) \int_{L_{NH}+L_H}^{z} ds \int_0^{L_{NH}} ds' A(s,s') \right.$$

$$+ \int_{L_{NH}+L_H}^{L_H+2L_{NH}} dz' \int_{L_{NH}+L_H}^{z} ds \int_{L_{NH}+L_H}^{z'} ds' A(s,s') + \int_0^{L_{NH}} dz' \int_0^{L_{NH}} ds \int_0^{z'} ds' A(s,s')$$

$$\left. +(L_H + L_{NH}) \int_0^{L_{NH}} ds \int_0^{L_{NH}} ds' A(s,s') + \int_{L_{NH}+L_H}^{2L_{NH}+L_H} dz' \int_0^{L_{NH}} ds \int_0^{z'} ds' A(s,s') \right] + \langle \chi^2 \rangle_{\partial\Omega},$$

(B.15)

where $A(s,s')$ is given by Eq. (B.11). We find that Eqs (B.13)-(B.15) become (using Eq. (3) of the main text)

$$\langle \Delta\tilde{\Phi}(z)^2 \rangle_{\partial\Omega} = \frac{2z^2}{\lambda_c^{(0)}(L_H + 2L_{NH})} - \frac{2z}{\lambda_c^{(0)}} + \frac{2L_{NH}(L_H + L_{NH})}{\lambda_c^{(0)}(L_H + 2L_{NH})} + \frac{4L_{NH}^3}{3\lambda_c^{(0)}(L_H + 2L_{NH})^2},$$

(B.16)

$$\langle \Delta\tilde{\Phi}(z)^2 \rangle_{\partial\Omega} = \frac{4L_{NH}^3}{3\lambda_c^{(0)}(L_H + 2L_{NH})^2},$$

(B.17)

$$\langle \Delta\tilde{\Phi}(z)^2 \rangle_{\partial\Omega} = \frac{4L_{NH}^3}{3\lambda_c^{(0)}(L_H + 2L_{NH})^2} + \frac{2(s - L_{NH} - L_H)^2}{(L_H + 2L_{NH})\lambda_c^{(0)}} + \frac{2L_H(s - L_{NH} - L_H)}{(L_H + 2L_{NH})\lambda_c^{(0)}}.$$

(B.18)

Let us consider again the integral $X_n$ defined in Eq. (A.29). Now, we write the sum

$$X_n = X_n^{(1)} + X_n^{(2)} + X_n^{(3)},$$

(B.19)

where

$$X_n^{(1)} \equiv \int_0^{L_{NH}} dz \exp\left(-\frac{n^2 \langle \Delta\tilde{\Phi}(z)^2 \rangle}{2}\right) = \sqrt{\frac{\lambda_c(L_H + 2L_{NH})\pi}{4n^2}} \exp\left(-\frac{n^2(8L_{NH}^3 - 3L_H^3 - 6L_{NH}L_H^2)}{12\lambda_c^{(0)}(L_H + 2L_{NH})^2}\right) \times$$

$$\left[ \text{erf}\left(\sqrt{\frac{n^2}{\lambda_c^{(0)}(L_H + 2L_{NH})}}\left(L_{NH} + \frac{L_H}{2}\right)\right) - \text{erf}\left(\sqrt{\frac{n^2}{\lambda_c^{(0)}(L_H + 2L_{NH})}}\left(\frac{L_H}{2}\right)\right) \right],$$

(B.20)

$$X_n^{(2)} \equiv \int_{L_{NH}}^{L_{NH}+L_H} dz \exp\left(-\frac{n^2 \langle \Delta\tilde{\Phi}(z)^2 \rangle}{2}\right) = L\exp\left(-\frac{2n^2 L_{NH}^3}{3\lambda_c^{(0)}(L_H+2L_{NH})^2}\right), \quad (B.21)$$

$$X_n^{(3)} \equiv \int_{L_{NH}+L_H}^{2L_{NH}+L_H} dz \exp\left(-\frac{\langle \Delta\tilde{\Phi}(z)^2 \rangle}{2}\right) = X_n^{(1)}. \quad (B.22)$$

The last equality is found via the variable change $z' = L_{NH} - z$ (the symmetry of the problem demands that this must be the case). Thus, the average interaction energy (Eq. (A.4)) becomes

$$\langle E_{int} \rangle_\Omega = La_0 - La_1 X_p(L_{NH}, L_H, \lambda_c^{(0)})\cos(\Delta\bar{\Phi}_0) + La_2 X_h(L_{NH}, L_H, \lambda_c^{(0)}/4)\cos(2\Delta\bar{\Phi}_0),$$

(B.23)

where

$$X_p(L_{NH}, L_H, \lambda_c^{(0)}) = \sqrt{\pi\lambda_c^{(0)}(L_H+2L_{NH})}\exp\left(-\frac{(8L_{NH}^3 - 3L_H^3 - 6L_{NH}L_H^2)}{12\lambda_c^{(0)}(L_H+2L_{NH})^2}\right) \times$$
$$\left[\mathrm{erf}\left(\sqrt{\frac{1}{(L_H+2L_{NH})\lambda_c^{(0)}}}\left(L_{NH}+\frac{L_H}{2}\right)\right) - \mathrm{erf}\left(\sqrt{\frac{1}{(L_H+2L_{NH})\lambda_c^{(0)}}}\left(\frac{L_H}{2}\right)\right)\right] + L\exp\left(-\frac{2L_{NH}^3}{3\lambda_c^{(0)}(L_H+2L_{NH})^2}\right).$$

(B.24)

### Imbedded Antiparallel Homologous Sequences

For the case of antiparallel homologous sequences, where, as before, $\delta\Omega_1(L_H + 2L_{NH} - z) = \delta\Omega_2(z)$ within the homology region $L_{NH} < z < L_H + L_{NH}$, and $\langle \delta\Omega_1(z)\delta\Omega_2(z') \rangle_{\delta\Omega} = 0$ elsewhere. Again, we start by finding $\langle \chi^2 \rangle_{\delta\Omega}$. From Eq. (B.5), this is given by the expression

$$\langle \chi^2 \rangle_{\delta\Omega} = \frac{1}{h^2(L_H+2L_{NH})^2}\Bigg[(L_H+L_{NH})^2 \int_0^{L_{NH}} ds \int_0^{L_{NH}} ds' A(s,s')$$
$$+ L_{NH}^2 \int_{L_{NH}}^{L_{NH}+L_H} ds \int_{L_{NH}}^{L_{NH}+L_H} ds' B(s,s') + 2(L_H+L_{NH})\int_0^{L_{NH}} dz' \int_0^{L_{NH}} ds \int_0^{z'} ds' A(s,s')$$
$$+ 2L_{NH}\int_{L_{NH}}^{L_{NH}+L_H} dz' \int_{L_{NH}}^{L_{NH}+L_H} ds \int_{L_{NH}}^{z'} ds' B(s,s') + \int_0^{L_{NH}} dz \int_0^{L_{NH}} dz' \int_0^{z} ds \int_0^{z'} ds' A(s,s')$$
$$+ \int_{L_{NH}}^{L_{NH}+L_H} dz \int_{L_{NH}}^{L_{NH}+L_H} dz' \int_{L_{NH}}^{z} ds \int_{L_{NH}}^{z'} ds' B(s,s') + \int_{L_{NH}+L_H}^{2L_{NH}+L_H} dz \int_{L_{NH}+L_H}^{2L_{NH}+L_H} dz' \int_{L_{NH}+L_H}^{z} ds \int_{L_{NH}+L_H}^{z'} ds' A(s,s')\Bigg],$$

(B.25)

where

$$B(s,s') = \langle \delta\Omega_1(s)\delta\Omega_1(s') \rangle_{\delta\Omega} + 2\langle \delta\Omega_1(L_H+2L_{NH}-s)\delta\Omega_1(s') \rangle_{\delta\Omega}$$
$$+ \langle \delta\Omega_1(L_H+2L_{NH}-s)\delta\Omega_1(L_H+2L_{NH}-s') \rangle_{\delta\Omega}.$$

(B.26)

Again using Eq. (3) of the main text we are able to evaluate Eq. (B.25), yielding

$$\left\langle \chi^2 \right\rangle_{\partial\Omega} = \frac{2L_{NH}(L_H + L_{NH})}{\lambda_c^{(0)}(L_H + 2L_{NH})} + \frac{4L_{NH}^3}{3\lambda_c^{(0)}(L_H + 2L_{NH})^2} + \frac{L_H^3}{3\lambda_c^{(0)}(L_H + 2L_{NH})^2}. \tag{B.27}$$

Now we have for $\left\langle \Delta\tilde{\Phi}(z)^2 \right\rangle$ the following expressions

$$\left\langle \Delta\tilde{\Phi}(z)^2 \right\rangle_{\partial\Omega} = \frac{1}{h^2}\int_0^z ds \int_0^z ds' A(s,s') - \frac{1}{h^2(L_H + 2L_{NH})}\int_0^{L_{NH}} dz' \int_0^z ds \int_0^{z'} ds' A(s,s')$$
$$-\frac{(L_H + L_{NH})}{h^2(L_H + 2L_{NH})}\int_0^z ds \int_0^{L_{NH}} ds' A(s,s') + \left\langle \chi^2 \right\rangle_{\partial\Omega},$$
when $0 < z \leq L_{NH}$,

(B.28)

$$\left\langle \Delta\tilde{\Phi}(z)^2 \right\rangle_{\partial\Omega} = \left\langle \Delta\tilde{\Phi}(L_{NH})^2 \right\rangle_{\partial\Omega} + \frac{1}{h^2}\int_{L_{NH}}^z ds \int_{L_{NH}}^z ds' B(s,s') - \frac{1}{h^2(L_H + 2L_{NH})}\int_{L_{NH}}^{L_{NH}+L_H} dz' \int_{L_{NH}}^z ds \int_{L_{NH}}^{z'} ds' B(s,s')$$
$$-\frac{L_H}{h^2(L_H + 2L_{NH})}\int_{L_{NH}}^z ds \int_{L_{NH}}^{L_{NH}+L_H} ds' B(s,s'),$$
when $L_{NH} < z \leq L_{NH} + L_H$,

(B.29)

$$\left\langle \Delta\tilde{\Phi}(z)^2 \right\rangle_{\partial\Omega} = \left\langle \Delta\tilde{\Phi}(L_{NH} + L_H)^2 \right\rangle_{\partial\Omega} + \frac{1}{h^2}\int_{L_H+L_{NH}}^z ds \int_{L_H+L_{NH}}^z ds' A(s,s')$$
$$-\frac{1}{h^2(L_H + 2L_{NH})}\int_{L_H+L_{NH}}^{2L_{NH}+L_H} dz' \int_{L_H+L_{NH}}^z ds \int_{L_H+L_{NH}}^{z'} ds' A(s,s'),$$
when $L_{NH} + L_H < z \leq 2L_{NH} + L_H$,

(B.30)

where $A(s, s')$ and $B(s, s')$ are given by Eqs. (B.11) and (B.26). Using Eq. (3) of the main text, Eqs. (B.28)-(B.30) become

$$\left\langle \Delta\tilde{\Phi}(z)^2 \right\rangle_{\partial\Omega} = \frac{2z^2}{\lambda_c^{(0)}(L_H + 2L_{NH})} - \frac{2s}{\lambda_c^{(0)}} + \left\langle \chi^2 \right\rangle_{\partial\Omega}, \qquad \text{for } 0 < z \leq L_{NH}, \tag{B.31}$$

$$\left\langle \Delta\tilde{\Phi}(z)^2 \right\rangle_{\partial\Omega} = \frac{2L_{NH}^2}{\lambda_c^{(0)}(L_H + 2L_{NH})} - \frac{2L_{NH}}{\lambda_c^{(0)}} + \frac{4(s - L_{NH})^2}{\lambda_c^{(0)}(L_H + 2L_{NH})} + \frac{2(2L_{NH} - L_H)(s - L_{NH})}{\lambda_c^{(0)}(L_H + 2L_{NH})}$$
$$-\frac{2\theta(2s - 2L_{NH} - L_H)}{\lambda_c^{(0)}}(2s - 2L_{NH} - L_H) + \left\langle \chi^2 \right\rangle_{\partial\Omega},$$

$$\text{for } L_{NH} < z \leq L_{NH} + L_H, \tag{B.32}$$

$$\langle \Delta\tilde{\Phi}(z)^2 \rangle_{\partial\Omega} = \frac{2(z - L_H - L_{NH})^2}{\lambda_c^{(0)}(L_H + 2L_{NH})} - \frac{2L_{NH}(L_{NH} + L_H)}{\lambda_c^{(0)}(L_H + 2L_{NH})} + \frac{2L_H(z - L_H - L_{NH})}{\lambda_c^{(0)}(L_H + 2L_{NH})} + \langle \chi^2 \rangle_{\partial\Omega},$$

$$\text{for } L_{NH} + L_H < z \leq 2L_{NH} + L_H. \tag{B.33}$$

Let us again consider the integral $X_n$ defined by Eq. (A.29). In this case we can write it as the sum

$$X_n = X_n^{(1)} + X_n^{(2)} + X_n^{(3)} + X_n^{(4)}, \tag{B.34}$$

where now

$$X_n^{(1)} \equiv \int_0^{L_{NH}} dz \exp\left(-\frac{n^2 \langle \Delta\tilde{\Phi}(z)^2 \rangle}{2}\right)$$

$$= \sqrt{\frac{\lambda_c^{(0)}(L_H + 2L_{NH})\pi}{4n^2}} \exp\left(\frac{n^2(L_H^3 + 6L_{NH}L_H^2 - 8L_{NH}^3)}{12\lambda_c^{(0)}(L_H + 2L_{NH})^2}\right) \tag{B.35}$$

$$\left[\text{erf}\left(\sqrt{\frac{n^2}{(L_H + 2L_{NH})\lambda_c^{(0)}}}\left(L_{NH} + \frac{L_H}{2}\right)\right) - \text{erf}\left(\sqrt{\frac{n^2}{(L_H + 2L_{NH})\lambda_c^{(0)}}}\left(\frac{L_H}{2}\right)\right)\right],$$

$$X_n^{(2)} \equiv \int_{L_{Nh}}^{L_{NH} + L_H/2} dz \exp\left(-\frac{n^2 \langle \Delta\tilde{\Phi}(z)^2 \rangle}{2}\right)$$

$$= \sqrt{\frac{\lambda_c^{(0)}(L_H + 2L_{NH})\pi}{8n^2}} \exp\left(\frac{n^2(-L_H^3 - 6L_H^2 L_{NH} - 12L_H L_{NH}^2 + 8L_{NH}^3)}{24\lambda_c^{(0)}(L_H + 2L_{NH})^2}\right) \tag{B.36}$$

$$\left[\text{erf}\left(\sqrt{\frac{2n^2}{(L_H + 2L_{NH})\lambda_c^{(0)}}}\left(\frac{L_{NH}}{2} + \frac{L_H}{4}\right)\right) - \text{erf}\left(\sqrt{\frac{2n^2}{(L_H + 2L_{NH})\lambda_c^{(0)}}}\left(\frac{L_{NH}}{2} - \frac{L_H}{4}\right)\right)\right],$$

$$X_n^{(3)} \equiv \int_{L_{NH} + L_H/2}^{L_{NH} + L_H} dz \exp\left(-\frac{n^2 \langle \Delta\tilde{\Phi}(z)^2 \rangle}{2}\right) = X_n^{(2)}, \tag{B.37}$$

and lastly

$$X_n^{(4)} \equiv \int_{L_{NH} + L_H}^{2L_{NH} + L_H} dz \exp\left(-\frac{n^2 \langle \Delta\tilde{\Phi}(z)^2 \rangle}{2}\right) = X_n^{(1)}. \tag{B.38}$$

The last equalities in both Eqs. (B.37) and (B.38) can be shown by making the integration variable changes $z' = L_H + L_{NH} - z$ and $z' = L_H + 2L_{NH} - z$, respectively. Thus, we obtain for the average energy (Eq. (A.4)) the expression

$$\langle E_{\text{int}} \rangle_\Omega = La_0 - La_1 X_{ap}\left(L_{NH}, L_H, \lambda_c^{(0)}\right)\cos\left(\Delta\bar{\Phi}_0\right) + La_2 X_{ap}\left(L_{NH}, L_H, \lambda_c^{(0)}/4\right)\cos\left(2\Delta\bar{\Phi}_0\right),$$

(B.39)

with

$$X_{ap}\left(L_{NH}, L_H, \lambda_c^{(0)}\right) = \sqrt{\pi \lambda_c^{(0)}(L_H + 2L_{NH})}\left[\exp\left(\frac{\left(L_H^3 + 6L_{NH}L_H^2 - 8L_{NH}^3\right)}{12\lambda_c^{(0)}\left(L_H + 2L_{NH}\right)^2}\right)\times\right.$$

$$\left[\text{Erf}\left(\sqrt{\frac{1}{\lambda_c^{(0)}(L_H + 2L_{NH})}}\left(L_{NH} + \frac{L_H}{2}\right)\right) - \text{Erf}\left(\sqrt{\frac{1}{\lambda_c^{(0)}(L_H + 2L_{NH})}}\left(\frac{L_H}{2}\right)\right)\right]$$

$$+ \exp\left(\frac{-L_H^3 - 6L_H^2 L_{NH} - 12L_H L_{NH}^2 + 8L_{NH}^3}{24\lambda_c^{(0)}(L_H + 2L_{NH})^2}\right)\frac{1}{\sqrt{2}}\times$$

$$\left.\left[\text{Erf}\left(\sqrt{\frac{2}{\lambda_c^{(0)}(L_H + 2L_{NH})}}\left(\frac{L_{NH}}{2} + \frac{L_H}{4}\right)\right) - \text{Erf}\left(\sqrt{\frac{2}{\lambda_c^{(0)}(L_H + 2L_{NH})}}\left(\frac{L_{NH}}{2} - \frac{L_H}{4}\right)\right)\right]\right].$$

(B.40)

The recognition energy for parallel homology is simply provided by subtracting Eq. (B.9) from Eq. (B.23), while the recognition energy for antiparallel homology is given by subtracting Eq. (B.9) from subtracted from Eq. (B.39).

## Appendix C. Flexible molecules

Here we consider the effect of the torsional flexibility of DNA. The total pairing energy is now described by Eqs. (10) and (4) of the main text. For the average energy described by Eq. (A.4), the averaging formula given by Eq. (A.5) applies. For flexible molecules we now use Eq. (11) of the main text to compute $\langle \Delta\tilde{\Phi}(z)^2 \rangle_{\partial\Omega}$, appearing in Eq. (A.5), where $\Delta\tilde{\Phi}(z) = \Delta\Phi(z) - \Delta\bar{\Phi}_0$. Here, we allow for $x \neq 0$, when the regions of homology of the pairs may not be directly overlapping, and consider each of the three types of fragment pairs in turn. First we deal with the uncorrelated sequence case.

### *Completely random pairs*

From Eqs. (11) and (3) we find that

$$\langle \Delta\tilde{\Phi}(z)^2 \rangle_{\partial\Omega} = \frac{\lambda_h}{2\lambda_c^{(0)}}.$$

(C.1)

For the torsional energy contribution (Eq. (1.10) of main text)

$$\left\langle \left(\frac{d\Delta\Phi(z)}{dz} - \frac{\delta\Omega_1(z) - \delta\Omega_2(z)}{h}\right)^2 \right\rangle_{\partial\Omega} = \frac{1}{2\lambda_c^{(0)}\lambda_h}.$$

(C.2)

Thus, we can write for the average pairing energy per unit length for random sequences

$$\varepsilon_{pair}^r = \frac{\langle E_{pair}\rangle_{\delta\Omega}}{L-x} = \frac{C}{8\lambda_h \lambda_c^{(0)}} + a_0(R) - a_1(R)\exp\left(-\frac{\lambda_h}{4\lambda_c^{(0)}}\right)\cos\Delta\bar{\Phi}_0 + a_2(R)\exp\left(-\frac{\lambda_h}{\lambda_c^{(0)}}\right)\cos 2\Delta\bar{\Phi}_0.$$

(C.3)

Minimization of Eq. (C.3) with respect to $\lambda_h$ yields the equation

$$\lambda_h = \sqrt{\frac{C}{2\left(a_1(R)\exp\left(-\frac{\lambda_h}{4\lambda_c^{(0)}}\right)\cos\Delta\bar{\Phi}_0 - 4a_2(R)\exp\left(-\frac{\lambda_h}{\lambda_c^{(0)}}\right)\cos 2\Delta\bar{\Phi}_0\right)}}.$$

(C.4)

Minimization of Eq. (C.13) yields either

$$\cos\Delta\bar{\Phi}_0 = \frac{a_1(R)}{4a_2(R)}\exp\left(\frac{3\lambda_h}{4\lambda_c^{(0)}}\right),$$

(C.5)

or

$$\cos\Delta\bar{\Phi}_0 = 1.$$

(C.6)

For the values of $a_1(R)$ and $a_2(R)$ considered in the main text, Eq. (C.6) applies.

### *Pairs with parallel homologous segments*

Upon considering the interaction between homologous pairs, as their homologous sections are shifted by a distance $x$, we may neglect finding the interaction energy between the pairs outside the region of homology. Thus, we suppose that for all $z$ the relationship $\Omega_1(z+x) = \Omega_2(z)$. The expression for $\langle\Delta\tilde{\Phi}(z)^2\rangle_{\delta\Omega}$ in this case becomes

$$\langle\Delta\tilde{\Phi}(z)^2\rangle_{\delta\Omega} = \frac{1}{4h^2}\int_{-\infty}^{\infty}dz'\int_{-\infty}^{\infty}dz''\,\text{sgn}(z-z')\text{sgn}(z-z'')\exp\left(-\frac{|z-z'|}{\lambda_h}\right)\exp\left(-\frac{|z-z''|}{\lambda_h}\right)C(z',z''),$$

(C.7)

where

$$C(z',z'') = \langle\delta\Omega_1(z')\delta\Omega_1(z'')\rangle_{\delta\Omega} - 2\langle\delta\Omega_1(z'+x)\delta\Omega_1(z'')\rangle_{\delta\Omega} + \langle\delta\Omega_1(z'+x)\delta\Omega_1(z''+x)\rangle_{\delta\Omega}$$
$$= \frac{2h^2}{\lambda_c^{(0)}}\left(\delta(z'-z'') + \delta(z'+x-z'')\right).$$

(C.8)

Evaluation of the integrals in Eq. (C.7) then yields

$$\langle\Delta\tilde{\Phi}(z)^2\rangle_{\delta\Omega} = \frac{\lambda_h}{2\lambda_c^{(0)}}\psi\left(\frac{|x|}{\lambda_h}\right), \qquad \text{with} \qquad \psi(y) = 1 - (1-y)\exp(-y).$$

(C.9)

The average torsional energy is given by

$$\left\langle \left( \frac{d\Delta\Phi(z)}{dz} - \frac{\delta\Omega_1(z) - \delta\Omega_2(z)}{h} \right)^2 \right\rangle_{\delta\Omega} = \frac{1}{4h^2\lambda_h^2} \int_{-\infty}^{\infty} dz' \int_{-\infty}^{\infty} dz'' \exp\left(-\frac{|z-z'|}{\lambda_h}\right) \exp\left(-\frac{|z-z''|}{\lambda_h}\right) C(z', z'').$$

(C.10)

Evaluating the integral in Eq. (C.10) yields

$$\left\langle \left( \frac{d\Delta\Phi(z)}{dz} - \frac{\delta\Omega_1(z) - \delta\Omega_2(z)}{h} \right)^2 \right\rangle_{\delta\Omega} = \frac{1}{2\lambda_h \lambda_c^{(0)}} \rho\left(\frac{|x|}{\lambda_h}\right),$$

(C.11)

where
$$\rho(y) = 1 - (1+y)\exp(-y).$$
(C.12)

Thus, we may find the average pairing energy per unit length (of the homology region)

$$\varepsilon_{pair}^p = \frac{C}{8\lambda_h \lambda_c^{(0)}} \rho\left(\frac{|x|}{\lambda_h}\right) + a_0(R)$$
$$-a_1(R)\exp\left(-\frac{\lambda_h}{4\lambda_c^{(0)}} \psi\left(\frac{|x|}{\lambda_h}\right)\right)\cos\Delta\bar\Phi_0 + a_2(R)\exp\left(-\frac{\lambda_h}{\lambda_c^{(0)}} \psi\left(\frac{|x|}{\lambda_h}\right)\right)\cos 2\Delta\bar\Phi_0.$$

(C.13)

Minimization of Eq. (C.13) with respect to $\lambda_h$ yields the equation

$$\lambda_h = \sqrt{\frac{C}{2\left(a_1(R)\exp\left(-\frac{\lambda_h}{4\lambda_c^{(0)}}\psi\left(\frac{|x|}{\lambda_h}\right)\right)\cos\Delta\bar\Phi_0 - 4a_2(R)\exp\left(-\frac{\lambda_h}{\lambda_c^{(0)}}\psi\left(\frac{|x|}{\lambda_h}\right)\right)\cos 2\Delta\bar\Phi_0\right)}}.$$

(C.14)

Minimization of Eq. (C.13) with respect to $\Delta\bar\Phi_0$ yields either

$$\cos\Delta\bar\Phi_0 = \frac{a_1(R)}{4a_2(R)}\exp\left(\frac{3\lambda_h}{4\lambda_c^{(0)}}\psi\left(\frac{|x|}{\lambda_h}\right)\right),$$

(C.15)

or $$\cos\Delta\bar\Phi_0 = 1.$$ (C.16)

Since, again, in the main text we consider only interaction coefficients where $a_1(R) > 4a_2(R)$, we only need consider the latter case, $\cos\Delta\bar\Phi_0 = 1$

*Pairs with antiparallel homologous segments*

Because the limits of integration of Eq. (1.11) of the main text are infinite, some care needs to be taken in carrying out the analysis for antiparallel homology. This situation, for the flexible case, is described by the choice is that $\Omega_1(-z-x) = \Omega_2(z)$ over the homology region, which is assumed to be infinite. Thus, we can write

$$\langle \Delta\tilde{\Phi}(z)^2 \rangle_{\partial\Omega} = \frac{1}{4h^2} \int_{-\infty}^{\infty} dz' \int_{-\infty}^{\infty} dz'' \, \mathrm{sgn}(z-z')\mathrm{sgn}(z-z'') \exp\left(-\frac{|z-z'|}{\lambda_h}\right) \exp\left(-\frac{|z-z''|}{\lambda_h}\right) D(z', z''),$$

(C.17)

where

$$D(z', z'') = \langle \delta\Omega_1(z')\delta\Omega_1(z'') \rangle_{\partial\Omega} - 2\langle \delta\Omega_1(-z'-x)\delta\Omega_1(z'') \rangle_{\partial\Omega} + \langle \delta\Omega_1(-z'-x)\delta\Omega_1(-z''-x) \rangle_{\partial\Omega}$$
$$= \frac{2h^2}{\lambda_c^{(0)}} \left( \delta(z'-z'') + \delta(z'+x+z'') \right).$$

(C.18)

Eq. (C.17) yields

$$\langle \Delta\tilde{\Phi}(z)^2 \rangle = \frac{\lambda_h}{2\lambda_c^{(0)}} \psi\left( \frac{|x+2z|}{\lambda_h} \right).$$

(C.19)

Upon taking the other average, Eq. (C.10), we find

$$\left\langle \left( \frac{d\Delta\Phi(z)}{dz} - \frac{\delta\Omega_1(z) - \delta\Omega_2(z)}{h} \right)^2 \right\rangle_{\partial\Omega} = \frac{1}{4h^2\lambda_h^2} \int_{-\infty}^{\infty} dz' \int_{-\infty}^{\infty} dz'' \exp\left(-\frac{|z-z'|}{\lambda_h}\right) \exp\left(-\frac{|z-z''|}{\lambda_h}\right) D(z', z'').$$

(C.20)

Eq. (C.20) becomes

$$\left\langle \left( \frac{d\Delta\Phi(z)}{dz} - \frac{\delta\Omega_1(z) - \delta\Omega_2(z)}{h} \right)^2 \right\rangle_{\partial\Omega} = \frac{1}{2\lambda_h\lambda_c^{(0)}} \rho\left( \frac{|x+2z|}{\lambda_h} \right).$$

(C.21)

Here, we may write the average energy per unit length (of the homology region) as

$$\varepsilon_{pair}^{ap} = a_0(R) - \frac{a_1(R)}{L_H} \int_{-L_H/2}^{L_H/2} dz \exp\left( -\frac{\lambda_h}{4\lambda_c^{(0)}} \psi\left( \frac{|x+2z|}{\lambda_h} \right) \right) \cos \Delta\bar{\Phi}_0$$
$$+ \frac{a_2(R)}{L_H} \int_{-L_H/2}^{L_H/2} dz \exp\left( -\frac{\lambda_h}{\lambda_c^{(0)}} \psi\left( \frac{|x+2z|}{\lambda_h} \right) \right) \cos 2\Delta\bar{\Phi}_0 + \frac{C}{8\lambda_h\lambda_c^{(0)}L_H} \int_{-L_H/2}^{L_H/2} dz \rho\left( \frac{|x+2z|}{\lambda_h} \right)$$
$$\approx \varepsilon_{pair}^{r} - \frac{a_1(R)}{L_H} \int_{-L_H/2}^{L_H/2} dz \sum_{m=1}^{\infty} \frac{(-1)^m}{m!} \left( \frac{\lambda_h}{4\lambda_c^{(0)}} \Delta\psi\left( \frac{|x+2z|}{\lambda_h} \right) \right)^m \cos \Delta\bar{\Phi}_0$$
$$+ \frac{a_2(R)}{L_H} \int_{-L_H/2}^{L_H/2} dz \sum_{m=1}^{\infty} \frac{(-1)^m}{m!} \left( \frac{\lambda_h}{\lambda_c^{(0)}} \Delta\psi\left( \frac{|x+2z|}{\lambda_h} \right) \right)^m \cos 2\Delta\bar{\Phi}_0 + \frac{C}{8\lambda_h\lambda_c^{(0)}L_H} \int_{-L_H/2}^{L_H/2} dz \Delta\rho\left( \frac{|x+2z+2L_{NH}|}{\lambda_h} \right)$$
$$\approx \varepsilon_{pair}^{r} + O(\lambda_h / L_H),$$

(C.22)

where $\Delta\rho(y) = -(1+y)\exp(-y)$ and $\Delta\psi(y) = -(1-y)\exp(-y)$. As our approximation, described by Eq. (11) of the main text, was originally for $L_H / \lambda_h \gg 1$ we should examine Eq. (C.22) in this limit. Here, the difference between $\varepsilon_{pair}^{ap}$ and $\varepsilon_{pair}^{r}$ becomes negligible. Thus, these analytical calculations predict no difference between the pairing energies, per unit length, of completely random, uncorrelated sequences and those pairs with antiparallel homologous segments. This is indeed seen in the simulation data. Unfortunately, we cannot say much about the size of the leading order correction in Eq. (C.22), as we have already neglected interface effects between the homologous and random regions of each fragment by making the choice $\Omega_1(-z-x) = \Omega_2(z)$ over the whole length, as well as finite size effects. The latter are neglected due to setting the limits of integration to infinity. However the simulation data has been fit with an empirical formula that relies on a leading order correction that has the form $\propto \lambda_h / L_H$, which this calculation suggests.

*The recognition energies*

In the limit $L_H / \lambda_h \gg 1$, the recognition energy for the case of parallel homologues is given by

$$E_{rec} = L_H \left( \varepsilon_{pair}^{p} - \varepsilon_{pair}^{r} \right), \tag{C.23}$$

whereas for the antiparallel case $E_{rec} \approx 0$, neglecting $O(\lambda_h)$ corrections.

*Including Thermal Torsional Fluctuations*

Here, we will only consider two cases, molecular pairs with random, uncorrelated sequences and those with parallel homologous sequences. It is fairly straight forward, using the previous arguments, to show that for the anti-parallel homologous case, again in the limit $L_H / \lambda_h \gg 1$, the pairing energy per unit length is nearly that for uncorrelated sequences. Let us first consider the general statistical mechanical problem, for which we do not yet need to consider either of these specific cases

The partition function for the pair is given by

$$Z_{pair} = \int D\Delta\Phi(z) \exp\left( -\frac{E_{pair}[\Delta\Phi(z)]}{k_B T} \right). \tag{C.24}$$

where the energy functional $E_{pair}[\Delta\Phi(z)]$ is given by Eqs. (4) and (10) of the main text.

To approximate the free energy we write a variational trial functional of the form

$$E_0 = \int_{-\infty}^{\infty} dz \left[ \frac{C}{4} \left( \frac{\delta\Phi(z)}{dz} \right)^2 + \frac{\alpha k_B T}{2} \delta\Phi(z)^2 \right], \tag{C.25}$$

where $\delta\Phi(z) = \Delta\Phi(z) - \langle\Delta\Phi(z)\rangle$, and here the averaging bracket corresponds to a thermal average. The variational free energy is then given by

$$F_T = k_B T \ln Z_0 + \langle E_{pair} - E_0 \rangle_0. \tag{C.26}$$

Next, we consider the thermal averages of the cosines:

$$\langle \cos n\Delta\Phi(z) \rangle_0 = \cos\left(n\langle\Delta\Phi(z)\rangle_0\right) \exp\left(-\frac{n^2 \langle \delta\Phi(z)^2 \rangle_0}{2}\right). \tag{C.27}$$

We find that $\langle \delta\Phi(z)^2 \rangle_0$ is given by

$$\langle \delta\Phi(z)^2 \rangle_0 = \frac{1}{2\pi} \int_{-\infty}^{\infty} dk \frac{1}{l_c k^2/2 + \alpha} = \frac{\lambda_h^T}{l_c}, \qquad \text{where} \quad \lambda_h^T = \left(\frac{l_c}{2\alpha}\right)^{1/2}, \tag{C.28}$$

where $l_c = C/k_B T$. For the variational trial function for $\langle \Delta\Phi(z) \rangle_0$ we find (similar to Eq. (11) for the ground state considered in the main text)

$$\langle \Delta\Phi(z) \rangle_0 = \Delta\bar{\Phi}_0 + \frac{1}{2h} \int_{-\infty}^{\infty} \text{sgn}(z - z') \exp\left(-\frac{|z - z'|}{\lambda_h}\right) [\delta\Omega_1(z') - \delta\Omega_2(z')] dz'. \tag{C.29}$$

Now let us consider the individual cases of pairing. Thus, using previous results (Eqs. (C.1) and (C.9)), we find that ensemble average of Eq. (C.27) for uncorrelated sequences is

$$\langle \langle \cos n\Delta\Phi(z) \rangle_0 \rangle_{\delta\Omega} = \cos n\Delta\bar{\Phi}_0 \exp\left(-n^2 \left(\frac{\lambda_h^T}{2l_c} + \frac{\lambda_h}{4\lambda_c^{(0)}}\right)\right), \tag{C.30}$$

whereas, in the case of pairs containing parallel homology,

$$\langle \langle \cos n\Delta\Phi(z) \rangle_0 \rangle_{\delta\Omega} = \cos n\Delta\bar{\Phi}_0 \exp\left(-n^2 \left(\frac{\lambda_h^T}{2l_c} + \frac{\lambda_h}{4\lambda_c^{(0)}} \psi\left(\frac{|x|}{\lambda_h}\right)\right)\right). \tag{C.31}$$

Also, taking into account torsional flexibility of the molecules, we need to consider the average

$$\left\langle \left(\frac{d\Delta\Phi(z)}{dz} - \frac{\delta\Omega_1(z) - \delta\Omega_2(z)}{h}\right)^2 \right\rangle_0 - \left\langle \left(\frac{d\delta\Phi(z)}{dz}\right)^2 \right\rangle_0 = \left(\frac{d\langle\Delta\Phi(z)\rangle_0}{dz} - \frac{\delta\Omega_1(z) - \delta\Omega_2(z)}{h}\right)^2.$$

$$\tag{C.32}$$

Performing the ensemble average on Eq. (C.32) for the uncorrelated sequences yields

$$\left\langle \left\langle \left(\frac{d\Delta\Phi(z)}{dz} - \frac{\delta\Omega_1(z) - \delta\Omega_2(z)}{h}\right)^2 \right\rangle_0 \right\rangle_{\delta\Omega} - \left\langle \left(\frac{d\delta\Phi(z)}{dz}\right)^2 \right\rangle_0 = \frac{1}{2\lambda_h \lambda_c^{(0)}}, \tag{C.33}$$

and for pairs containing parallel homology

$$\left\langle \left\langle \left( \frac{d\Delta\Phi(z)}{dz} - \frac{\delta\Omega_1(z) - \delta\Omega_2(z)}{h} \right)^2 \right\rangle_0 \right\rangle_{\delta\Omega} - \left\langle \left( \frac{d\delta\Phi(z)}{dz} \right)^2 \right\rangle_0 = \frac{1}{2\lambda_h \lambda_c^{(0)}} \rho\left( \frac{|x|}{\lambda_h} \right). \tag{C.34}$$

Upon replacing the infinite limits of integration in Eq.(C.25) with $L_H/2$ and $-L_H/2$ and using the fact that

$$\frac{\partial \ln Z_0}{\partial \alpha} = -\frac{L_H}{2} \left\langle \delta\Phi(z)^2 \right\rangle, \tag{C.35}$$

we find that the contribution from the remaining terms in the free energy is (up to an unimportant constant)

$$-\frac{1}{L_H} \ln Z_0 + \frac{\alpha}{2} \left\langle \delta\Phi(z)^2 \right\rangle = \frac{1}{4\lambda_h^T}. \tag{C.36}$$

Thus, we find that the reduced free energy per unit length for completely random molecules, $f_{rand} = F_T / (L_H k_B T)$ is given by

$$f_{rand} = \frac{1}{4\lambda_h^T} + \frac{l_c}{8\lambda_c^{(0)} \lambda_h} + \frac{a_0(R)}{k_B T} - \frac{a_1(R)}{k_B T} \exp\left( -\left( \frac{\lambda_h^T}{2l_c} + \frac{\lambda_h}{4\lambda_c^{(0)}} \right) \right)$$
$$+ \frac{a_2(R)}{k_B T} \exp\left( -\left( \frac{2\lambda_h^T}{l_c} + \frac{\lambda_h}{\lambda_c^{(0)}} \right) \right). \tag{C.37}$$

In the case of the molecules with embedded homologous sequences the free energy per unit length (of the homology region), for $f_p = F_T / (L_H k_B T)$, is given by

$$f_p = \frac{1}{4\lambda_h^T} + \frac{l_c}{8\lambda_c^{(0)} \lambda_h} \rho\left( \frac{|x|}{\lambda_h} \right) + \frac{a_0(R)}{k_B T} - \frac{a_1(R)}{k_B T} \exp\left( -\left( \frac{\lambda_h^T}{2l_c} + \frac{\lambda_h}{4\lambda_c^{(0)}} \psi\left( \frac{|x|}{\lambda_h} \right) \right) \right)$$
$$+ \frac{a_2(R)}{k_B T} \exp\left( -\left( \frac{2\lambda_h^T}{l_c} + \frac{\lambda_h}{\lambda_c^{(0)}} \psi\left( \frac{|x|}{\lambda_h} \right) \right) \right). \tag{C.38}$$

In both cases, we find $\lambda_h^T = \lambda_h$. Also, we find that $\lambda_h$ satisfies, for uncorrelated sequences, the expression

$$\lambda_h = \sqrt{\frac{C}{2\left( a_1(R) \cos \Delta\bar{\Phi}_0 \exp\left( -\lambda_h \left( \frac{1}{2l_c} + \frac{1}{4\lambda_c^{(0)}} \right) \right) - 4a_2(R) \cos 2\Delta\bar{\Phi}_0 \exp\left( -\lambda_h \left( \frac{2}{l_c} + \frac{1}{\lambda_c^{(0)}} \right) \right) \right)}},$$
(C.39)

and for pairs containing parallel homology

$$\lambda_h = \left( \frac{2}{C} \left( a_1(R) \cos \Delta\bar{\Phi}_0 \exp\left( -\lambda_h \left( \frac{1}{2l_c} + \frac{1}{4\lambda_c^{(0)}} \psi\left( \frac{|x|}{\lambda_h} \right) \right) \right) \right. \right.$$

$$\left. \left. -4a_2(R) \cos 2\Delta\bar{\Phi}_0 \exp\left( -\lambda_h \left( \frac{2}{l_c} + \frac{1}{\lambda_c^{(0)}} \psi\left( \frac{|x|}{\lambda_h} \right) \right) \right) \right) \right)^{-1/2}.$$

(C.40)

The minimum values of $\Delta\bar{\Phi}_0$ are either given by $\cos \Delta\bar{\Phi}_0 = 1$, or for uncorrelated sequences by

$$\cos \Delta\bar{\Phi}_0 = \frac{a_1(R)}{4a_2(R)} \exp\left( 3\lambda_h \left( \frac{1}{2l_c} + \frac{1}{4\lambda_c^{(0)}} \right) \right),$$

(C.41)

and for pairs containing parallel homology by

$$\cos \Delta\bar{\Phi}_0 = \frac{a_1(R)}{4a_2(R)} \exp\left( 3\lambda_h \left( \frac{1}{2l_c} + \frac{1}{4\lambda_c^{(0)}} \psi\left( \frac{|x|}{\lambda_h} \right) \right) \right).$$

(C.42)

# Appendix D. Numerical results with $L_{NH} \neq 0$

Here, we show simulation results for the recognition energy for both rigid and flexible molecules as a function of $L_{NH}$. The values $a_1 = 0.015 k_B T / \text{Å}$ and $a_2 = 0.0035 k_B T / \text{Å}$ are used, as in the main text.

In Fig D.1 we present the simulation results for rigid molecules. We find that when the averaging of the interaction energy over the realizations of $\delta\Omega_1(z) - \delta\Omega_2(z)$ is performed before optimizing $\Delta\Phi_0$, there is very little change in the recognition energy values. The only change that we can discern is a slight increase in the magnitude of the recognition energy for fragments containing parallel homology. This is effect presumably due to exponential factors that appear in the analytical result Eq. (A.37). These factors are suppressed when $L_{NH} \gg \lambda_c^{(0)} = 150\text{Å}$, and at these values of $L_{NH}$ we would indeed expect the value of the recognition energy to be almost constant.

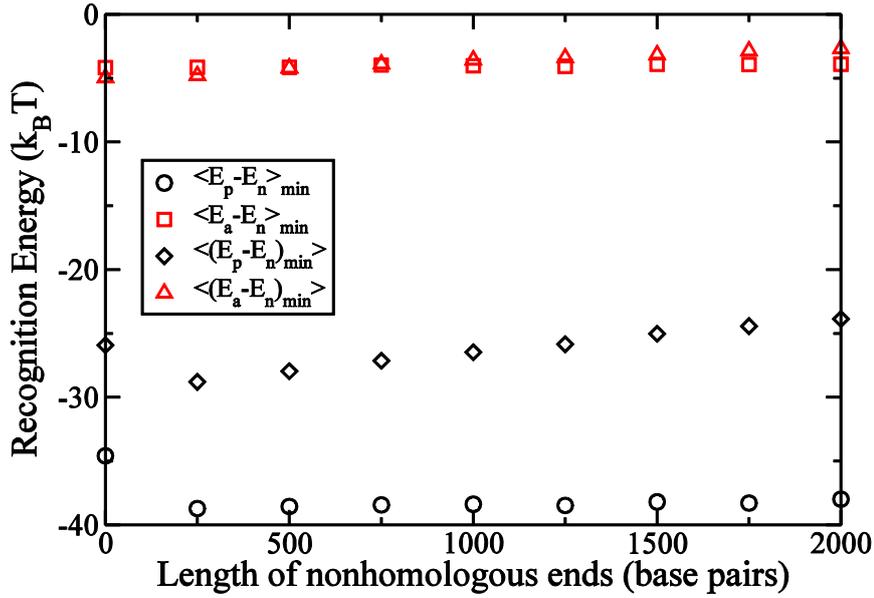

Fig. D.1. The simulation data for the recognition energy (at $x=0$) with respect to $L_{NH}$; the energy difference between fragment pairs containing homology and those completely uncorrelated. The simulation is done for rigid fragments ($C=\infty$) calculated with a homology length $L_H = 1000$Å. The black points correspond to pairs with parallel orientated homology, whereas the red ones correspond to antiparallel homology. Both the circles and squares correspond optimizing the relative azimuthal orientations of the fragments after calculating the average ensemble energy. The triangles and diamonds correspond to optimizing the azimuthal orientations for each base pair configuration before averaging the energy over realizations of base pair dependent helix distortions.

Now let us consider changing the order of averaging for rigid molecules (c.f. Fig D.1). In other words, we optimize $\Delta\Phi_0$ for each realization of $\delta\Omega_1(z) - \delta\Omega_2(z)$; i.e. the value of $\Delta\Phi_0$ that satisfies Eq. (B.1). Here, we do indeed see a slight decrease in the magnitude of the recognition energy with increasing $L_{NH}$. This decrease is predicted by perturbation theory, but for large values of $L_H$ and $L_{NH}$ the perturbation theory is quantitatively inaccurate, hence we do not show in the plots the results for such values. Note that the perturbation theory predicts a far more rapid reduction in the recognition energy magnitude than what is seen in the simulation results [61].

Next, in Fig. D.2 we present simulation results for torsionally flexible molecules for the recognition energy per homology length $E_{rec}/L_H$, with $L_H = 1000$Å. We see that slight changes that were seen for rigid molecules now disappear, as the order of averaging no longer matters.

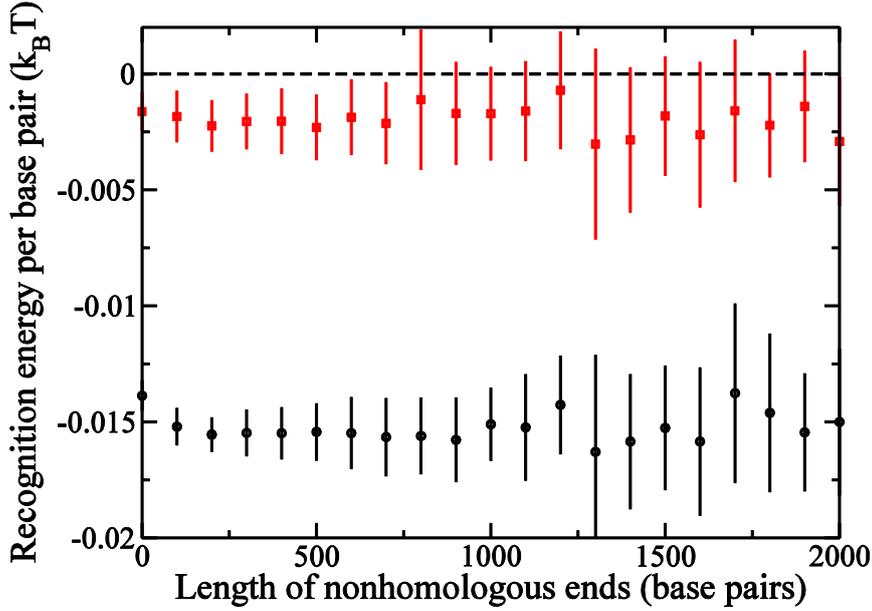

Fig. D.2. The simulation data for the recognition energy per base pair of homology (at $x=0$), given by $E_{rec}/L_H$, with respect to $L_H$. Here a homology length of $L_H = 1000\text{Å}$ is used, and a value of torsional rigidity $C/k_B T = 725\text{Å}$. The red points corresponds to anti-parallel homology and the black points to parallel homology.

## Appendix E. Numerical results for finite temperature

Here we use the finite temperature results given by Eqs. (C.37)-(C.40) where we have again used the parameter values $a_1 = 0.015 k_B T/\text{Å}$ and $a_2 = 0.0035 k_B T/\text{Å}$. For these values we may set $\Delta\bar{\Phi} = 0$. In Fig. E.1. we compare these results with the case where thermal fluctuations are not considered; the analytical approximations for both completely uncorrelated pairs and those containing parallel homology are given by Eqs. (C.3), (C.4), (C.13) and (C.14). We find that the effect of finite temperature on the recognition energy is very slight, and actually deepens the well by a small amount, for the value $L_{NH} = 6800\text{Å}$. Presumably, this effect is due the magnitude of the pairing energy of uncorrelated sequences being affected more by thermal fluctuations.

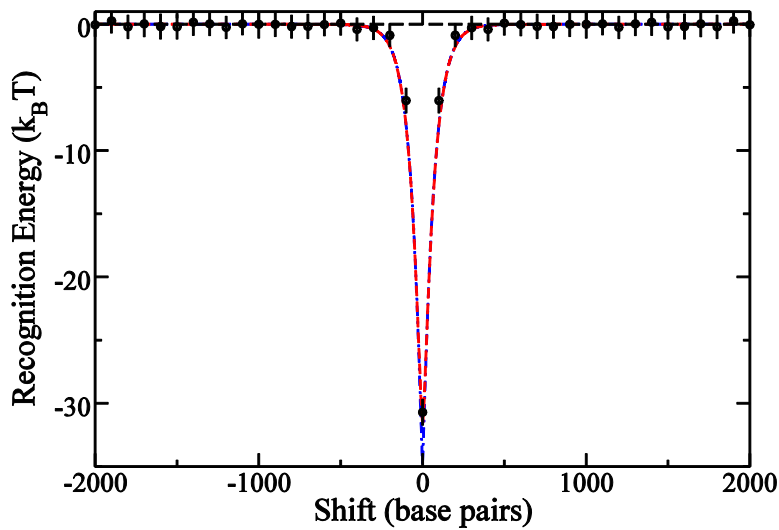

Fig. E.1. Comparison of the recognition well including thermal fluctuations with that calculated for the ground-state. The plots of the recognition well, calculated using the analytical approximations, are shown for the ground state (shown in red) and with thermal fluctuations (shown in blue). For completeness the (ground state) simulation data is given by black points. In these plots a homology length of $L_H = 6800\text{Å}$ (corresponding to $2000$ base pairs) was used, and $L_{NH} = 0$.

[29] A. Leforestier, F. Livolant, Proc. Natl. Acad. Sci. USA, **106**, 9157 (2009).

[30] A .A. Kornyshev, D. J. Lee, S. Leikin, A. Wynveen, S. Zimmerman, Phys. Rev. Lett., **95**, 148102 (2005).

[31] A. Wynveen, D. J. Lee, A. Kornyshev, S. Leikin, Nucl. Acids Res., **36**, 5540 (2008)

[32] A. A. Kornyshev, S. Leikin, Phys.Rev.Lett., **86**, 3666(2001).

[33] A. G. Cherstvy, A. A. Kornyshev, S. Leikin, J. Phys. Chem.B, **108**, 6508 (2004).

[34] A. A. Kornyshev, A. Wynveen, Phys.Rev.E, **69**, 041905 (2004).

[35] R. Cortini, A. A. Kornyshev, D. J. Lee, S. Leikin, Biophys.J., **101**, 875 (2011)

[36] A. A. Kornyshev, S. Leikin, Phys. Rev. Lett., **82,** 4138 (1999)

[37] S. Inoe, S. Sugiyama, A. A. Travers, T. Ohyama, Biochemistry **46**, 164 (2007).

[38] G. S. Baldwin, N. J. Brooks, R. E. Robson, A. Wynveen, A. Goldar, S. Leikin, J. M. Seddon, A. A. Kornyshev, J. Phys. Chem. B, **112**, 1060 (2008).

[39] J. Nishikawa, T. Ohyama, *Nucleic Acid Res. 41,* 1544 (2013).

[40] C. Danilovicz, C. H. Lee, K. Kim, K. Hatch, V. W. Coljee, N. Kleckner, M. Prentiss, Proc. Nat. Acad. Sci. USA **106**, 19824 (2009).

[41] X. Wang, X. Zhang, C. Mao, N. C. Seeman, Proc. Nat .Acad. Sci. USA **107**, 12547 (2010).

[42] A. A. Kornyshev, A. Wynveen, Proc. Natl. Acad. Sci. USA*,* **106**, 4683 (2009).

[43] D. J. Lee, A. A. Kornyshev, J. Chem. Phys **131**,155104 (2009); erratum: J. Chem. Phys. **131**, 219901 (2009).

[44] Within the same organisms, there is only a tiny difference at most between the identical DNA texts and homologous ones (the same type of genes, responsible for the same encoding function).

[45] M. E. Hood, M. Katawcik, T. Giraud, Genetics, **170**, 1081 (2005).

[46] E. Gladyshev, N. Kleckner, Nat. Commun., **5**, 3509 (2014)

[47] A. A. Kornyshev, D. J. Lee, S. Leikin, A. Wynveen, Nucl. Acids Res. **39**, 7289 (2011)

[48] As $L_{NH}$, the length of the non-homologous sections, increases, the perturbation theory suggests that the net value of the recognition energy decreases. This is different from the recognition energy per base pair, which always decreases with increasing $L_{NH}$, no matter which way we do the averaging. The former effect occurs because the nonhomologous sections will tend to find their optimal azimuthal alignments which may not allow to align optimally the homologous section. This can be seen from Eq. (9), where the optimum phase $\Delta\Phi_0$ depends on the spatial average of the accumulated mismatch, the integral of $\delta\Omega_1(z') - \delta\Omega_2(z')$. For $L_{NH} << L_H$, $\Delta\Phi_0$ lies close to the value that minimizes the energy for the homologous segments, and we would expect a significant recognition energy. However, as $L_{NH}$ increases, $\Delta\Phi_0$ will change to a value closer to one that minimizes the energy of the uncorrelated sequence segments, away from the optimum alignment for the homologous segments; therefore, the magnitude of the recognition energy should decrease. This prediction is confirmed in the MC simulations, although this happens at a much slower rate than suggested by the perturbation theory (for the plots see Appendix D of the Supplemental Material).

[49] D. J. Lee, http://arxiv.org/abs/1312.1568

[50] In the mean field electrostatic model of charged interacting helices [A. A. Kornyshev, S. Leikin, J. Chem Phys **107,** 3656 (1997)], these correspond to the parameter values $\theta = 0.85$, $f_1 = 0.35$ and $f_2 = 0.65$ (for details and formulas see either [A. A. Kornyshev, S. Leikin, J. Chem Phys **107**, 3656 (1997)] or [A. A. Kornyshev, D. J. Lee, S. Leikin, A. Wynveen, Rev. Mod. Phys., **79**, 943 (2007)]) with inter-axial separation $R = 30\text{Å}$.

[51] This value represents the mid-point of experimental values presented in [A. A. Kornyshev, D. J. Lee, S. Leikin, A. Wynveen, Rev. Mod. Phys., **79**, 943 (2007)]. In later works we have sometimes used a higher value of $C/k_B T = 1000\text{Å}$, reported in [J. Lipfert, J. W. J. Kerssemakers, T. Jager, N. H. Dekker, Nature Methods, **7**, 977 (2010) ]. This different choice does not change much the results presented here.

[52] A. Noy, R.Golestanian, Phys. Rev. Lett. , **109**, 228101 (2012).

[53] This result has some implication for experiments similar to those of Baldwin et al [G. S. Baldwin, N. J. Brooks, R. E. Robson, A. Wynveen, A. Goldar, S. Leikin, J. M. Seddon, A. A. Kornyshev, J. Phys. Chem. B, **112**, 1060 (2008)]. In these experiments two different sequences of DNA were used, labelled by different colour fluorophores, was observed for fragments of 298 base pairs, within cholesteric spherulites formed at mild osmotic stress. In this case, and for shorter fragments, the DNA can be approximated as rigid. Due to the

nature of the labelling, these segregation experiments cannot distinguish between anti-parallel and parallel alignments. Based on our results, we might reasonably expect some phase segregation could happen for fragments with less than 250 base pairs. However each region, containing only one of the types if base pair sequence, could consist of either mixtures of anti-parallel and parallel aligned homologous molecules or smaller domains of parallel aligned sequences orientated in either two directions. The study of this situation poses an interesting problem in statistical mechanics. Though, to detect these alignments requires different experiments.

[54] D. J. Lee, A. Wynveen, J. Phys. Conds Matter, **18,** 787 (2006).

[55] The recognition of some small identical base tracks is known happen in site specific recombination processes [N. D. F. Grindley, K. L. Whiteson,. P. A. Rice, Ann. Rev. Biochem. **75**, 567 (2006))]. This recognition is facilitated by recognition proteins that recognize a particular base pair sequence.

[56] M. Prentiss, Private communication (2014).

[57] D. J. Lee, http://arxiv.org/abs/1411.4844

[58] R. Cortini, D. J. Lee, A. Kornyshev, J. Phys. Cond. Matter, **24,** 162203 (2012).

[59] J. F. Marko, E. D. Siggia, Phys. Rev. E. **52**, 2912 (1995).

[60] J. Ubbink, T. Odijk, Biophys. J. **76,** 2502 (1999).

[61] An expansion in powers of $\sin\left(\Delta\tilde{\Phi}(z)\right)$ might form the basis of a better analytical approximation for the solution to Eq. (B.1), but this has yet to be considered.